%% file: main.tex
\documentclass[10pt,journal,compsoc]{IEEEtran}

\usepackage[normalem]{ulem} % added for easing the review with strikethrough fonts
\usepackage{color} % added for easing the review with text colors
\usepackage[hyphens]{url} % added for url line breaking in footnotes
\usepackage{graphicx} 
\usepackage{amsmath} 
\usepackage{listings}
\usepackage{subfig}

\let\labelindent\relax
\usepackage{enumitem}

\begin{document}

% carregon: force to show 1st author + et al in references.
\bstctlcite{IEEEexample:BSTcontrol}

%\title{Intra-node Scheduling of Applications on GPUs}
\title{Intra-node Memory Safe GPU Co-Scheduling}

\author{Carlos Rea\~no, %~\IEEEmembership{Member,~IEEE,}
        Federico Silla, %~\IEEEmembership{Member,~IEEE}
        Dimitrios S. Nikolopoulos  %~\IEEEmembership{Member,~IEEE}
		and
        Blesson Varghese  %~\IEEEmembership{Member,~IEEE}

\IEEEcompsocitemizethanks{ \IEEEcompsocthanksitem 
C. Rea\~no and F. Silla are with the Universitat Polit\`ecnica de Val\`encia, Spain. D. Nikolopoulos and B. Varghese are with the Queen's University Belfast, UK. Email: {\tt\footnotesize carregon@gap.upv.es, fsilla@upv.es, \{d.nikolopoulos, varghese\}@qub.ac.uk}
}
\thanks{}}

\markboth{IEEE Transactions on Parallel and Distributed Systems, Accepted 12~December~2017}%
{1 \MakeLowercase{\textit{Rea\~no et al.}}: Title of paper}

\input{abstract}

\maketitle

\IEEEdisplaynotcompsoctitleabstractindextext
\IEEEpeerreviewmaketitle

\section{Introduction}
\label{introduction}
\input{introduction}

\section{GPU Scheduling Framework}
\label{framework}
\input{framework}

\section{Implementation Approaches}
\label{implementation}
\input{implementation}

\section{The Life Cycle}
\label{apifunctions}
\input{apifunctions}

\section{Notification Policies}
\label{clientnotification}
\input{policies}

\section{Experimental Setup}
\label{setup}
\input{setup}

\section{Evaluation}
\label{evaluation}
\input{evaluation}

\section{Related Work}
\label{relatedwork}
\input{relatedwork}

\section{Conclusions \textcolor{black}{and Future Work}}
\label{conclusions}
\input{conclusions}

%\ifCLASSOPTIONcompsoc
%   \section*{Acknowledgments}
% \else
   \section*{Acknowledgment}
% \fi
 \input{acknowledgment}

\ifCLASSOPTIONcaptionsoff
  \newpage
\fi

\bibliographystyle{IEEEtran}
\bibliography{references}
%\begin{thebibliography}{00}

%\end{thebibliography}

\begin{IEEEbiography}
[{\includegraphics[width=1in,height=1.25in,clip,keepaspectratio]{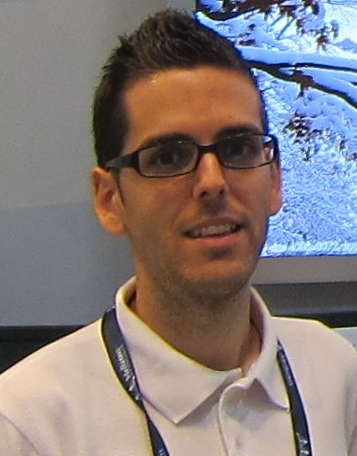}}]
{Carlos Rea\~no}
received the MSc (2012) and PhD (2017) degrees in Computer Engineering from Technical University of Valencia, Spain. He is currently a postdoctoral researcher at the Department of Computer Engineering of that university, focusing his research on the virtualisation of scarce resources, mainly GPUs. More information is available from \url{http://mural.uv.es/caregon}.
%received a BS degree in Computer Engineering from the University of Valencia, Spain, in 2008. He holds a MS degree in Software Engineering, Formal Methods and Information Systems since 2012 and is currently a PhD candidate at the Technical University of Valencia, Spain. He has published several papers in peer-reviewed conferences and journals and also contributes as a reviewer to these.
\end{IEEEbiography}

\vspace{-0.5cm}

\begin{IEEEbiography}
[{\includegraphics[width=1in,height=1.25in,clip,keepaspectratio]{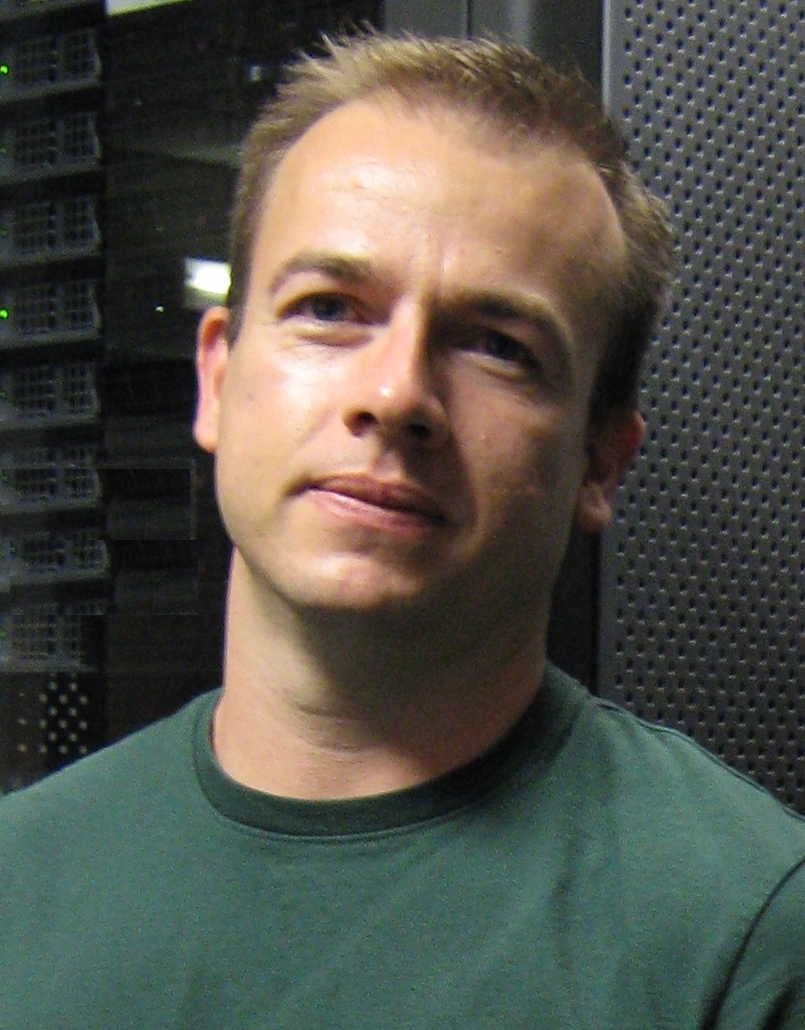}}]
{Federico Silla}
received the MS and PhD degrees from Technical University of Valencia, Spain. He is currently an associate professor at the Department of Computer Engineering of that university.
%, and since January 2016 is credited by the Spanish Government as Full Professor.
He has previously worked for the Intel Corporation. His research addresses high performance on-chip and off-chip interconnection networks, distributed memory systems and remote GPU virtualisation mechanisms. %Currently, he is coordinating the rCUDA remote GPU virtualization project since it began in 2008.
More information is available from \url{http://www.disca.upv.es/fsilla}.
\end{IEEEbiography}

\vspace{-0.5cm}

\begin{IEEEbiography}
[{\includegraphics[width=1in,height=1.25in,clip,keepaspectratio]{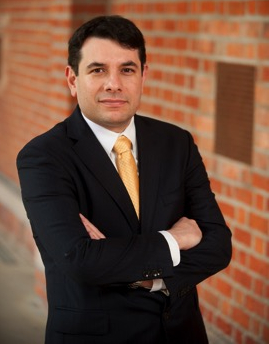}}]
{Dimitrios S. Nikolopoulos}
is Professor and Head of the School of Electronics, Electrical Engineering and Computer Science, at Queen's University of Belfast. He holds the Chair in High Performance and Distributed Computing. His research explores scalable computing systems for data-driven applications and new computing paradigms at the limits of performance, power and reliability. More information is available from \url{http://www.cs.qub.ac.uk/~D.Nikolopoulos/}.
%Dimitrios's accolades include the Royal Society Wolfson Research Merit Award, the NSF CAREER Award, the DOE CAREER Award, the IBM Faculty Award, the SFI-DEL Investigator Award and Best Paper Awards from the best IEEE and ACM HPC conferences, including SC, PPoPP, and IPDPS. His research has produced over 170 top-tier outputs and has received extensive (£11.1m as PI/£43.4m as CoI) and highly competitive research funding from the NSF, DOE, EPSRC, SFI, DEL, Royal Academy of Engineering, Royal Society, European Commission and private sector. He regularly teaches modules in computer organisation, parallel computing, and systems programming. Dimitrios is a Fellow of the British Computer Society, Senior Member of the IEEE and Senior Member of the ACM. He earned a PhD (2000) in Computer Engineering and Informatics from the University of Patras.
\end{IEEEbiography}

\vspace{-0.5cm}

\begin{IEEEbiography}
[{\includegraphics[width=1in,height=1.25in,clip,keepaspectratio]{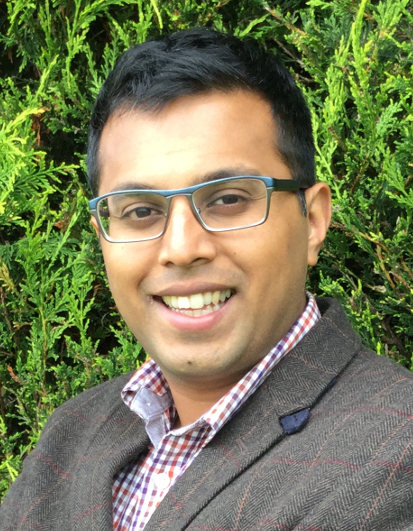}}]
{Blesson Varghese}
is a Lecturer at Queen’s University Belfast and an Honorary Lecturer at the University of St Andrews. He obtained a PhD in Computer Science (2011) and MSc in Network Centred Computing (2008), both from the University of Reading, UK, on international scholarships. Blesson's interests are in developing and analysing state-of-the-art parallel and distributed systems. More information is available from \url{www.blessonv.com}.
\end{IEEEbiography}

\end{document}

%% file: abstract.tex
\IEEEcompsoctitleabstractindextext{
\begin{abstract}
GPUs in High-Performance Computing systems remain under-utilised due to the unavailability of schedulers that can safely schedule multiple applications to share the same GPU. The research reported in this paper is motivated to improve the utilisation of GPUs by proposing a framework, we refer to as schedGPU, to facilitate intra-node GPU co-scheduling such that a GPU can be safely shared among multiple applications by taking memory constraints into account. Two approaches, namely a client-server and a shared memory approach are explored. However, the shared memory approach is more suitable due to lower overheads when compared to the former approach. 
Four policies are proposed in schedGPU to handle applications that are waiting to access the GPU, two of which account for priorities. The feasibility of schedGPU is validated on three real-world applications. The key observation is that a performance gain is achieved. 
For single applications, a gain of over 10 times, as measured by GPU utilisation and GPU memory utilisation, is obtained. For workloads comprising multiple applications, a speed-up of up to 5x in the total execution time is noted. Moreover, the average GPU utilisation and average GPU memory utilisation is increased by 5 and 12 times, respectively.

\end{abstract}

\begin{keywords}
GPU Co-scheduling, access synchronisation, memory safe, accelerator, under-utilisation, schedGPU
\end{keywords}}

%% file: introduction.tex
High-Performance Computing (HPC) systems are becoming heterogeneous in pursuit of exascale computing~\cite{gpucluster-1,gpucluster-3}.%~\cite{gpucluster-1,gpucluster-2,gpucluster-3}. 
These systems not only offer CPUs, but also provide accelerators, such as Graphics Processing Units (GPUs).
%carregon-v3\textcolor{red}{\sout{For example, Tianhe-2 and Titan listed among the top three supercomputers in the June 2016 Top500 list use accelerators}}\footnote{\textcolor{red}{\sout{https://www.top500.org/list/2016/06/}}}.  
Heterogeneity can be leveraged for improving performance by decomposing compute-intensive components of an application and offloading them on to GPUs. Existing schedulers, such as Slurm~\cite{slurm} and Torque~\cite{Torque}, cannot safely schedule multiple applications for sharing the same GPU~\cite{schedulerproblem-1}, thereby exclusively locking a GPU for a single application. This results in the under-utilisation of GPUs and will have negative implications on the performance of future exascale systems~\cite{gpuunderutilisation-1,gpuunderutilisation-2}.
%~\cite{gpuunderutilisation-1,gpuunderutilisation-2,gpuunderutilisation-3}. %It has also been reported, for example, that over 50\% of workloads on the Titan supercomputers do not use GPUs\footnote{\url{https://www.nextplatform.com/2015/08/11/lessons-learned-on-the-largest-gpu-supercomputer/}} and for a most part remain under-utilised. 
Hence, the research in this paper aims to improve the utilisation of GPUs. 

\begin{figure}
\begin{center}
\subfloat[Using existing workload schedulers]
{\label{fig:typical_scenario}
\includegraphics[width=0.43\textwidth]
{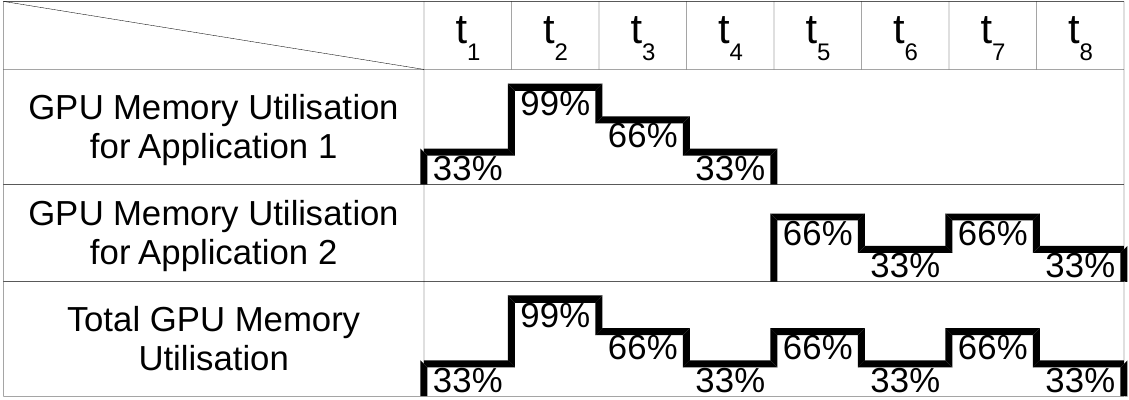}}

\subfloat[Using proposed approach]
{\label{fig:desired_scenario}
\includegraphics[width=0.43\textwidth]
{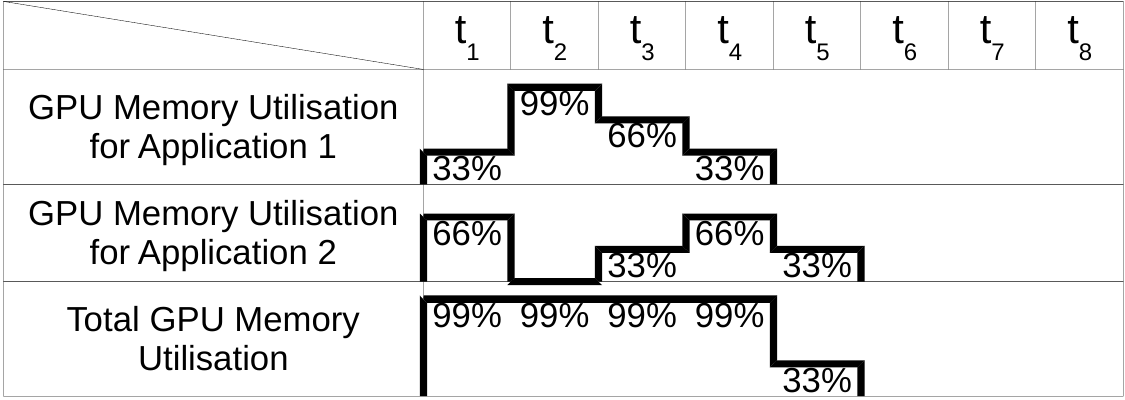}}
\end{center}
\caption{Execution of applications on a two CPU node with one GPU. In Figure~\ref{fig:typical_scenario}, two applications need access to the GPU, but are executed sequentially using existing workload schedulers. Figure~\ref{fig:desired_scenario} shows the proposed approach that co-schedules applications on the same GPU and accounts for GPU memory to maximise utilisation.}
\vspace{-0.5cm}
\label{fig:scenarios}
\end{figure}

One way of addressing accelerator under-utilisation, given that GPUs are usually coupled to a CPU node, is by sharing GPUs among multiple applications that execute on different CPU cores of the same node. However, there is a risk of running out of GPU memory which can cause applications to unexpectedly end. Current techniques that incorporate scheduling~\cite{SchedulingRef1,SchedulingRef3,SchedulingRef4}, kernel-based~\cite{GLoop, GPUSync, TimeGraph}, synchronisation~\cite{SchedulingRef6} and architectural~\cite{CUDA-MPS-Paper, CUDA-HyperQ-Paper} approaches cannot safely share GPUs among applications while eliminating the above risk.
Therefore, applications are currently executed sequentially, although they may use the GPU for a relatively small fraction of the entire execution time, as shown in Figure~\ref{fig:typical_scenario}.
This raises the need for a scheduler that can account for GPU memory required by applications to safely share GPUs as indicated in Figure~\ref{fig:desired_scenario}. 

In this paper, we propose an intra-node, memory safe GPU co-scheduling framework, referred to as \textit{schedGPU}.
The framework safely handles multiple application requests to access GPUs by ensuring that memory overruns do not occur during execution.
%\textcolor{red}{\sout{One advantage of using schedGPU is that minimum modifications need to be made to an application by using three simple functions offered by the framework.}} 
Two implementation approaches, namely a client-server and a shared memory approach, are considered. %The shared memory approach has fewer overheads than the client-server approach and is therefore suitable. 
The access of applications to shared memory is synchronised by developing a custom protocol that employs file locks and system signals. This protocol avoids abandonment, the problem that arises when the framework employs other interprocess synchronisation mechanisms, such as mutexes. Four policies are proposed and investigated in schedGPU to handle applications that wait to be scheduled on the GPU. %These policies also take priorities into account. 

The feasibility of schedGPU is validated \textcolor{black}{first using popular GPU benchmark suites and then} on three real-world applications that have varying GPU utilisation. Using schedGPU performance gain in terms of average speed-up, average GPU utilisation and average GPU memory utilisation
when executing concurrent instances of an application
using schedGPU is noted to be up to 10 times over conventional execution. For workloads comprising multiple applications, using Slurm along with schedGPU results in a 5x speed-up. Moreover, the average GPU utilisation and average GPU memory utilisation is increased by 5 and 12 times, respectively, when compared to not using schedGPU.

The research contribution of this paper is an approach for intra-node scheduling \textcolor{black}{at runtime}. Conventional schedulers schedule applications ahead-of-time typically over multiple nodes. However, they do not optimise scheduling on each node. 
The merit of our approach is that scheduling is performed on the fine-grain level \textcolor{black}{at runtime}, therefore allowing any application to be executed without knowledge of its GPU requirements prior to execution. 
%The proposed approach is GPU memory safe and co-schedules multiple applications. 
Existing schedulers exclusively lock a given GPU for an application. Our novel approach on the other hand is memory safe and shares GPUs to co-schedule multiple applications. For this our approach monitors the GPU resources to service applications.
%carregon-v3\textcolor{red}{\sout{The advantage of our approach is observed in terms of speed-up, GPU utilisation and GPU memory utilisation when executing applications. For example, a performance gain of up to 10 times is observed.}} 

The remainder of this paper is organised as follows.
Section~\ref{framework} presents the key concepts of our GPU co-scheduling framework.
Section~\ref{implementation} provides the implementation approaches considered in our framework. 
Section~\ref{apifunctions} describes the typical life cycle of an application using our framework. 
Section~\ref{clientnotification} details a set of four policies incorporated in the framework for scheduling applications. 
Section~\ref{setup} identifies suitable real-world use-cases for schedGPU.
\textcolor{black}{Section~\ref{evaluation} evaluates the performance of the framework using two popular GPU benchmark suites and then highlights its benefit for three real-world applications.}
Section~\ref{relatedwork} considers the related research.
Section~\ref{conclusions} concludes this paper \textcolor{black}{and presents future work}.

%% file: framework.tex
Consider a typical server, which comprises multiple CPU cores and one GPU. There are two challenges in executing multiple applications on the same GPU. Firstly, consider a scenario in which a conventional workload scheduler, such as Slurm, is employed to schedule applications from multiple users, then the scheduler will handle multiple requests to the GPU by simply executing the jobs sequentially. While these schedulers can schedule applications on multiple servers they schedule them ahead-of-time, leaving no room for adapting to \textcolor{black}{runtime} requirements. Therefore, the jobs are executed sequentially on each server since the scheduler cannot ensure whether the GPU memory requirements of requesting application can be met at any time (for example, whether there is sufficient GPU memory for a second application on the GPU). 

Secondly, assume a workload scheduler can schedule multiple applications on the same GPU. While this can improve GPU utilisation, it could lead to potentially terminating the job (for example, if there is insufficient GPU memory an out of memory error will be returned). This is because there is no safe handling of GPU memory requirements for co-scheduling jobs.

In this paper, we address the above challenges by presenting a framework for intra-node GPU scheduling, referred to as schedGPU\footnote{The schedGPU framework can be requested for download from \url{http://mural.uv.es/caregon/schedgpu.html}}, that facilitates the simultaneous execution of multiple applications on a GPU. Using schedGPU multiple applications can request GPU memory during execution time. schedGPU safely co-schedules the applications by taking memory requirements into account and thereby avoids potential memory allocation errors due to unavailable memory on the GPU. The schedGPU framework is proposed and developed for CUDA-based~\cite{CUDA} GPU applications. CUDA is widely used in production and commercial environments when compared to the OpenCL alternative~\cite{OpenCL}.

The features of the GPU scheduling framework are: 

(i) \textit{Intra-node scheduling}: most schedulers schedule applications over multiple nodes at the coarse-grain level. However, schedGPU schedules at the fine-grain level to improve GPU utilisation of each node.

(ii) \textit{Scheduling \textcolor{black}{at runtime}}: unlike conventional schedulers that schedule applications ahead-of-time, our framework can schedule applications on to the GPU in sub-millisecond timescales during execution. 

(iii) \textit{Memory-based safe co-scheduling}: typically schedulers allow for executing an application on multiple GPUs. Our approach facilitates the execution of multiple applications on the same GPU concurrently to improve GPU utilisation. We consider memory requirements of each application and ensure that no application unexpectedly ends due to insufficient GPU resources. 

%\item[iv] \textcolor{red}{\sout{\textit{Simplicity}: existing approaches for co-scheduling are more complex. For example, requires kernel modification as presented in Section~\ref{relatedwork}. However, the schedGPU approach cooperates with the application code by requiring a few additional lines of code, but does not require any system-level modifications.}}

(iv) \textit{Scalability}: the control of most workload schedulers is centralised. On the contrary, the control of schedGPU is distributed on each node hence avoiding single points of failure and use on a large number of nodes.   

An application that needs to safely use a GPU through our proposed framework follows a four stage life cycle. The first stage is initialising an instance of schedGPU for the application to allow interaction between the application and the framework. The second stage is reserving GPU memory required by an application, we refer to as pre-allocation. The GPU memory requests made by the application are appropriately handled by the framework. The third stage is releasing reserved GPU memory after the application makes use of the GPU, we refer to as post-free. Applications still waiting for the GPU are potentially serviced. The fourth stage is shutting down the instance of schedGPU that was initialised. 

%% file: implementation.tex
The schedGPU framework incorporates two approaches, namely a client-server and a shared memory model\footnote{In this paper, `shared memory' does not refer to aggregating host and device memory for offering a unified address space. This is because schedGPU schedules the access of applications to the GPU from the host side. Instead, we refer to shared memory as the host memory which is accessible for different applications using schedGPU. We refer to `shared memory data structure' as the format of the data and its contents that is stored in the shared memory.}.
\textcolor{black}{The latter is the focus of this paper.}
A prototype of the client-server approach was briefly reported elsewhere~\cite{ReanoHPCS2016}. %\textcolor{red}{\sout{The prototype, in contrast to the research reported in this paper, was not a generic framework for GPU applications, but was specific to an options pricing algorithm. The framework proposed and presented in this paper is a generic framework that can be used with any CUDA application. Additionally, the prototype did not share GPUs between applications at the same time and the research question was limited to increasing throughput of the specific algorithm thereby only provided restricted co-scheduling.}} 
%\textcolor{red}{\sout{However, in this paper, a GPU is shared between multiple applications simultaneously to improve the overall system performance measured by metrics relevant to modern servers.}}

%\textcolor{red}{\sout{The focus in this paper is the shared memory approach. The implementation of the client-server approach is similar to the shared memory approach in a number of ways. For example, the data stored by the shared memory structure is similar to that stored by the server. The access to the data in the server is synchronised using intraprocess mutexes and conditions, which are conceptually similar to interprocess mutexes employed in the shared memory approach. The life cycle of an application is similar to the client-server, with the exception that the initialisation and the shutdown stages are distributed between the client and the server. Hence, there are client and server initialisation and shutdown stages. Additionally, the policies for notifying waiting applications are not approach specific.}}
\textcolor{black}{The functionality of both the client-server and shared memory implementations is the same, which is to avoid applications running out of GPU memory. However, there are important differences architecturally. On one hand, the first approach follows a client-server architecture and the GPU information is centralised on the server side. On the other hand, the second approach performs a distributed management of the GPU information using shared memory.}

\subsection{Client-Server}
\label{clientserver}
In this approach, each CUDA application is a client that requests GPU memory to a centralised server daemon, both of which are executed on the same node. The server permits the application to continue execution if there is sufficient memory on the GPU, otherwise the client may choose to be either blocked until memory is available or informed using CUDA error codes. %Check the preceding statement

\begin{figure}
\centering
\includegraphics[width=0.36\textwidth]{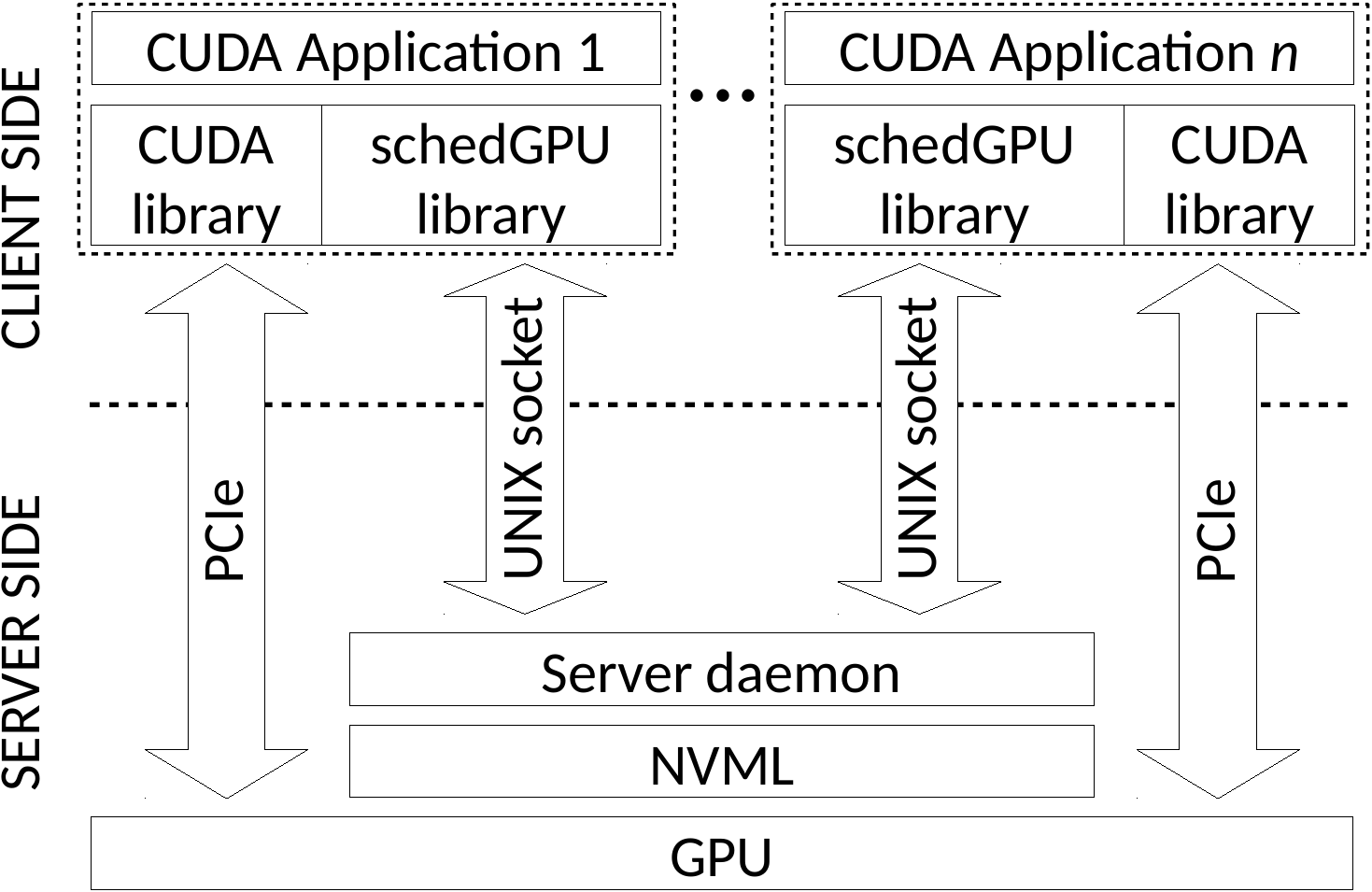}
\caption{Architecture of the client-server model}
\vspace{-0.5cm}
\label{fig:client_server_arch}
\end{figure}

Figure \ref{fig:client_server_arch} presents the architecture of the client-server model. In this model, the CUDA application is minimally modified by explicitly calling functions from the client library to pre-allocate the GPU memory required by the application. The calls are forwarded to the server using a UNIX domain socket. We chose UNIX domain sockets over TCP Loopback sockets due to the superior performance of the former \cite{unixsockets}. The server creates a new thread for each client. A global view of the memory used by all clients is maintained by the server through the NVIDIA Management Library (NVML)\footnote{\url{https://developer.nvidia.com/nvidia-management-library-nvml}}. We chose to use NVML instead of the CUDA library to avoid the creation of an additional GPU context that consumes GPU memory. 
In addition, using NVML the physical devices are accessed instead of using logical devices to avoid any ambiguities in the framework (for example, applications using different identifiers for logical devices referring to the same physical device). 

%Architectural Diagram

\subsection{Shared Memory}
\label{sharedmemory}
\begin{figure}
\centering
\includegraphics[width=0.36\textwidth]{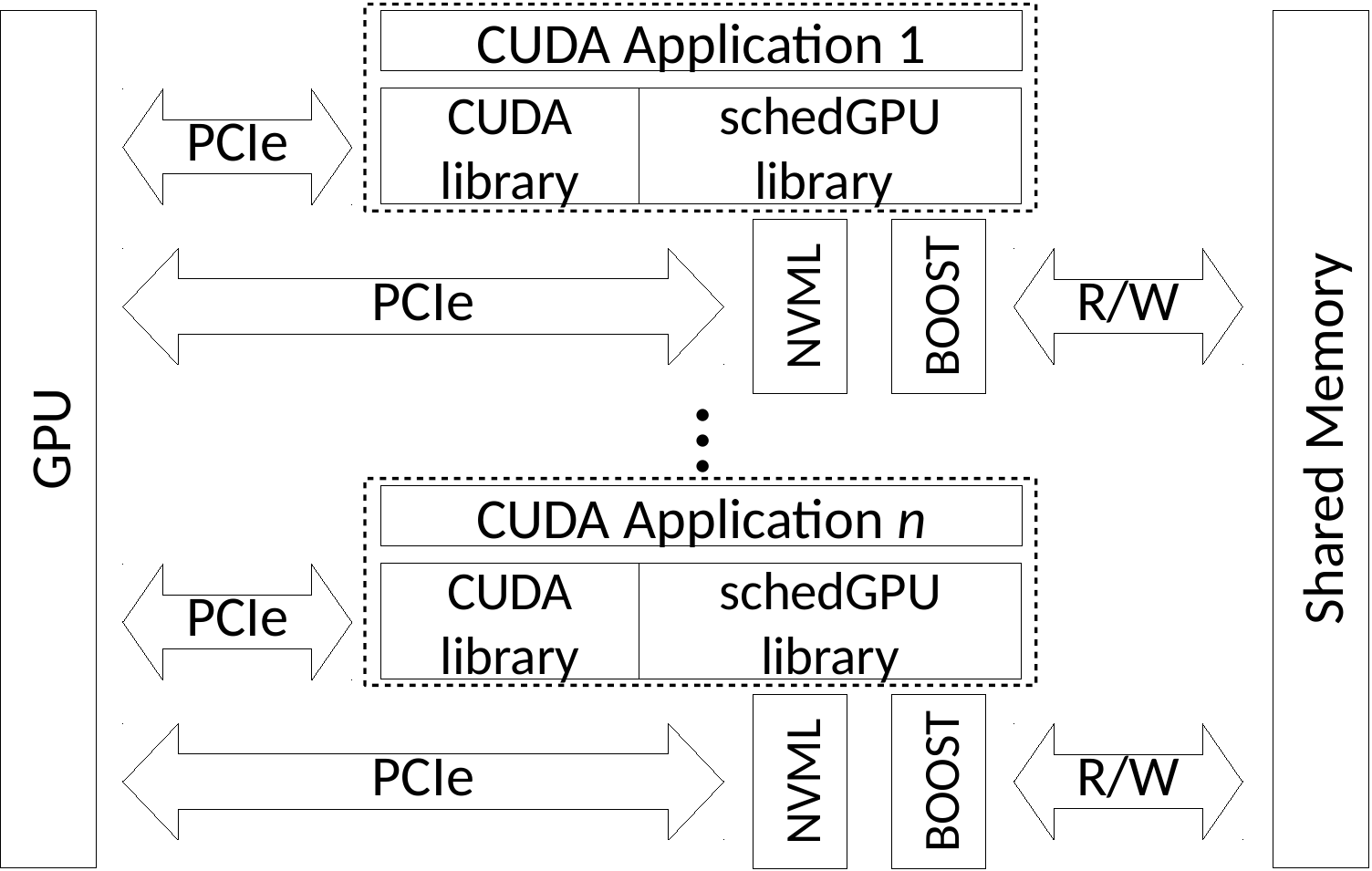}
\caption{Architecture of the shared memory model}
\vspace{-0.25cm}
\label{fig:shared_memory_arch}
\end{figure}

One disadvantage of a centralised server is that it may unexpectedly end resulting in the failure of the framework. Therefore, an alternate distributed approach was considered using shared memory in which the clients are responsible for maintaining the global state of the GPU memory in use. 
The client library makes decisions based on the information available in the shared memory data structure. Figure~\ref{fig:shared_memory_arch} shows the architecture of the shared memory model. The client library directly makes use of NVML. The shared memory structure is created and managed using the Boost Interprocess library\footnote{\url{http://www.boost.org/doc/libs/1_59_0_b1/doc/html/interprocess.html}}. 

The shared memory approach not only overcomes the disadvantages of the client-server approach, but also achieves better performance that will be considered in Section \ref{evaluation}. Therefore, this paper will focus on schedGPU using the shared memory approach. Four features are incorporated in the shared memory approach to enhance the robustness of the model. Firstly checkpointing by storing a backup of the shared memory structure by each client library when the application ends.

Secondly, an integrity check and recovery. When a new client starts, the library checks whether the shared memory structure is corrupt. If corrupt, then it is recovered from a backup. If there are no backups or if corrupt, then the shared memory structure is freshly initialised. 

Thirdly, a sanity check. When any client has ended or is blocked and waiting for free memory, the client library checks that processes with allocated memory are still alive (this is done to free the memory of clients that unexpectedly terminate). If not, the previously allocated memory is freed for the waiting clients.

Fourthly, mitigating abandonment. If a client application unexpectedly ends, the access to the shared memory structure is not blocked and the framework could be used by other clients transparently. If the client had memory allocated, it will be freed. 

%Architectural Diagram

%\subsubsection{Shared Memory Data}
%\label{shared_memory_contents}
\textit{Shared Memory Data}: 
The data structure used in the shared memory approach comprises the following data:
\begin{itemize}[leftmargin=0.3cm]
\item Total GPU memory: the total installed or physical device memory accessible to the schedGPU framework.
\item Total Used memory: the total memory utilised by active schedGPU client applications. 
\item Itemised Used memory: memory utilised by each client application that is uniquely identified by schedGPU. 
\item Queue of client applications waiting to access GPU memory: a queue of applications that requested more GPU memory than what was available. The priority of an application is also included depending on the policy in use. Policies will be described in Section~\ref{clientnotification}.
\end{itemize}

%\subsubsection{Synchronising Access to Shared Memory}
%\label{synchronisingaccess}
\textit{Synchronising Access to Shared Memory:}
To avoid inconsistencies we synchronise the access of multiple clients to the shared memory data structure using two methods.  

%\subsubsubsection{Interprocess Mutexes and Conditions}
%\label{interprocess_mutexes_and_conditions}
The first method is based on using interprocess mutexes and conditions both provided by the same Boost Interprocess Library that manages the shared memory data structure. Access of each client application to the shared memory data is controlled by a mutex which ensures that only one client modifies the shared memory structure at a given time. 
A condition is associated with the mutex for either notifying clients that memory has been freed or waiting for notifications on freed memory.

Although this is the most common method, it is restricted due to \textit{abandonment}: if a process owning the mutex unexpectedly ends, then the mutex becomes unusable and other processes endlessly wait for it. This can be avoided by the use of lock-free methods such as robust mutexes. Such methods are currently available for intraprocess communication (multi-threads). In the case of schedGPU, multiple applications will need to communicate with schedGPU simultaneously and this additionally requires interprocess communication along with intraprocess communication. However, there are no standard solutions for lock-free synchronisation in interprocess communication\footnote{\url{http://www.boost.org/doc/libs/1_61_0/boost/interprocess/detail/robust_emulation.hpp}}. One available solution requires patching the operating system kernel\footnote{\url{http://yurovsky.github.io/2015/06/04/lockfree-ipc/}}.

%\subsubsection{File Locks and System Signals}
In order to surmount the challenge of abandonment when using interprocess communication in schedGPU, we developed a second method that employs file locks instead of interprocess mutexes for controlling the access of client applications to the shared memory data structure. If a client owning a file lock unexpectedly ends, then the file lock is still safely used by other clients. 

However, conditions cannot be associated with file locks. So a custom protocol using system signals %\footnote{\url{http://man7.org/linux/man-pages/man7/signal.7.html}} 
was developed for interprocess communication. The protocol issues a user system signal for notifying a waiting client that memory has been freed. The waiting client on receiving the signal continues execution on the GPU. This method is more appropriate than the first method and is therefore incorporated in the schedGPU framework.

%% file: apifunctions.tex
Three functions are offered by schedGPU %\footnote{The API reference can be accessed from \url{http://mural.uv.es/caregon/schedgpu.html}} 
that implements the life cycle. They include the initialisation, memory pre-allocation and memory post-freeing functions. The shutdown stage in the framework is implicitly called when the CUDA application terminates execution. \textcolor{black}{schedGPU provides two options for using the functions: (i) implicit memory management - users do not modify the source code, but memory is implicitly managed by the framework using a set of default parameters, and (ii) explicit memory management - users minimally modify the GPU source code by including the schedGPU functions. This provides the developer finer control on memory management.}

\begin{figure}
\centering
\includegraphics[width=0.4\textwidth]{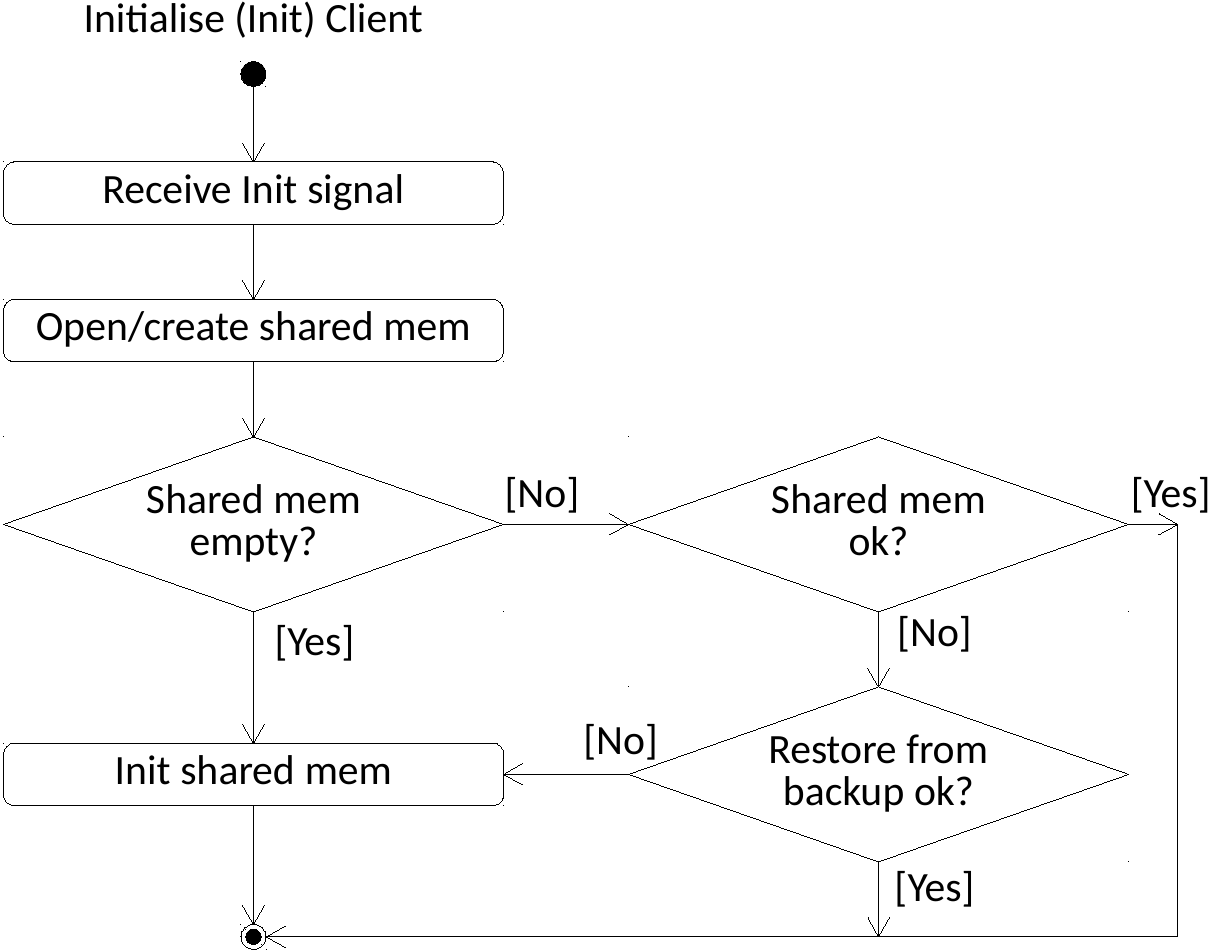}
\caption{Initialisation of a schedGPU client}
\vspace{-0.25cm}
\label{fig:shared_memory_activity_init}
\end{figure}

%\subsection{Initialisation}
\textit{Initialisation}: Figure~\ref{fig:shared_memory_activity_init} shows how a schedGPU client is initialised in the shared memory approach using the \texttt{schedGPUInit()} function. First of all, the client is made ready to handle system signals that are used for internal notifications. The shared memory data structure is then accessed or created. If the shared memory structure is empty, then it is initialised by gathering information of the GPUs using NVML. If the shared memory structure already contains GPU information, then an integrity check is performed to ensure that the data is not corrupted (recovers from the shared memory backup if the integrity check fails). 

%The \texttt{schedGPUInit()} function to initialise the scheduler must be the first function used in the CUDA program. The parameter \texttt{timeout\_in\_seconds} indicates the maximum time the scheduler should block the application when requesting GPU memory until control is given back to the CUDA program. The parameter \texttt{priority} indicates the priority of the application when requesting GPU memory. A higher \texttt{priority} value means a low priority.

%\begin{lstlisting}
%cudaError_t schedGPUInit(int timeout_in_seconds = 60, int priority = 0);
%\end{lstlisting}

%Error handling is implemented using the following standard CUDA error codes:
%\begin{itemize}[leftmargin=0.3cm]
%\item \texttt{cudaSuccess}: indicates that the scheduler was successfully initialised
% \item \texttt{cudaErrorUnknown}: notifies that the scheduler initialisation failed
%\end{itemize}

\begin{figure}
\centering
\includegraphics[width=0.42\textwidth]{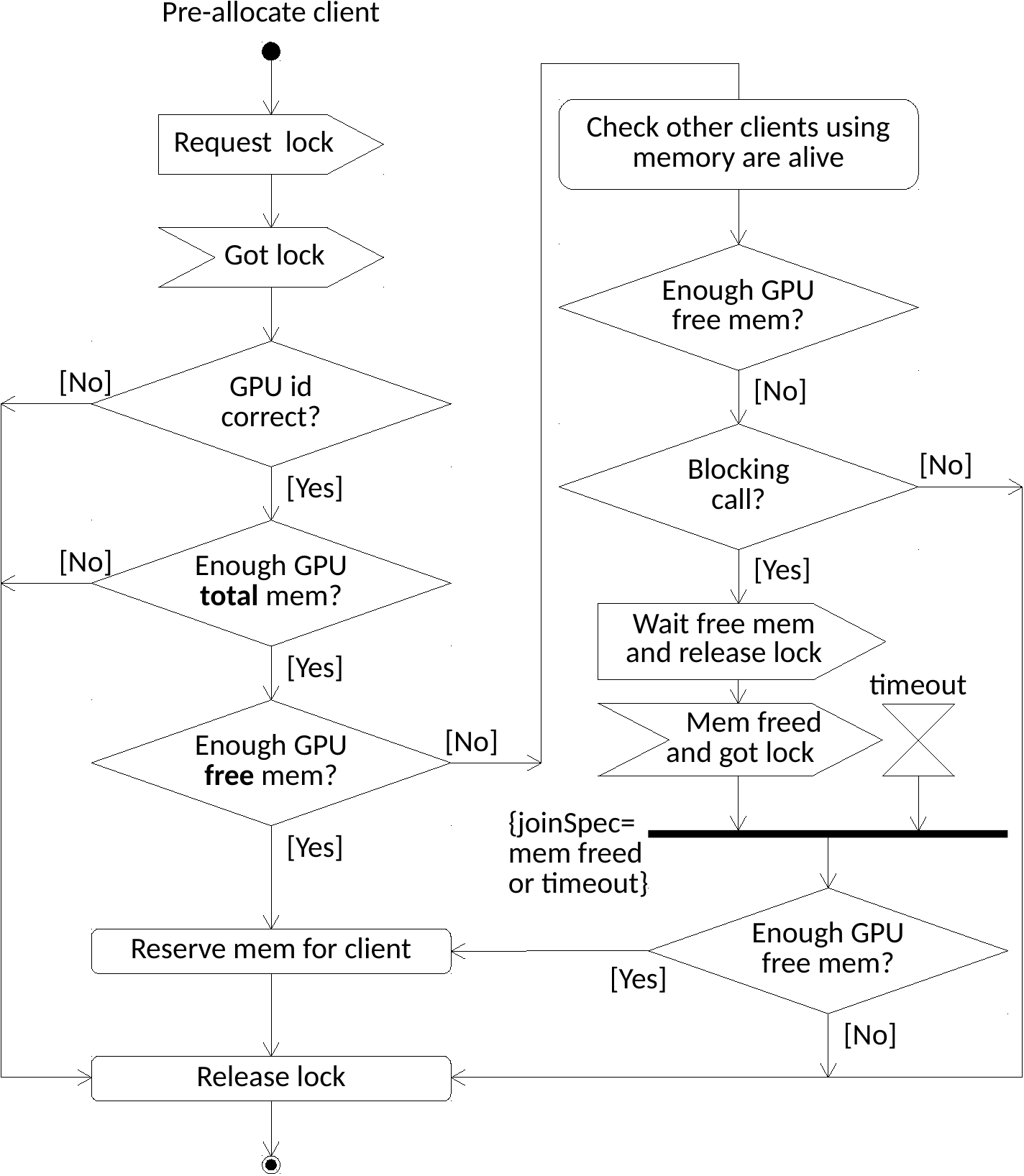}
\caption{\textcolor{black}{Pre-allocation of memory by a schedGPU client}}
\vspace{-0.35cm}
\label{fig:shared_memory_activity_prealloc}
\end{figure}

%\subsection{Pre-allocation}
\textit{Pre-allocation}: As shown in Figure~\ref{fig:shared_memory_activity_prealloc} 
for pre-allocating memory using the \texttt{preCudaMalloc()} function, the client requests the ownership of the file lock. When the ownership of the lock is obtained, the client checks that the GPU requested is a valid device and the requested memory is available on the GPU. If memory is insufficient, then the client performs a sanity check on whether other clients with pre-allocated memory are still alive. If memory is freed from other clients, then the client re-checks if there is sufficient memory.

If the available memory is still insufficient and provided that the pre-allocation call is non-blocking, then control is returned to the application with an error code (\texttt{cudaErrorNotReady}). If the call is blocking, then the client waits for a specified time period\textcolor{black}{\footnote{\textcolor{black}{This avoids a deadlock; the user either has to provide a timeout for the pre-allocation call or receives an instantaneous error message.}}} defined by the application until free memory is available. In the event that the client does not pre-allocate all free GPU memory, then it does not notify other clients of free memory. This notification is carried out during post-freeing. 

\begin{figure}
\centering
\includegraphics[width=0.3\textwidth]{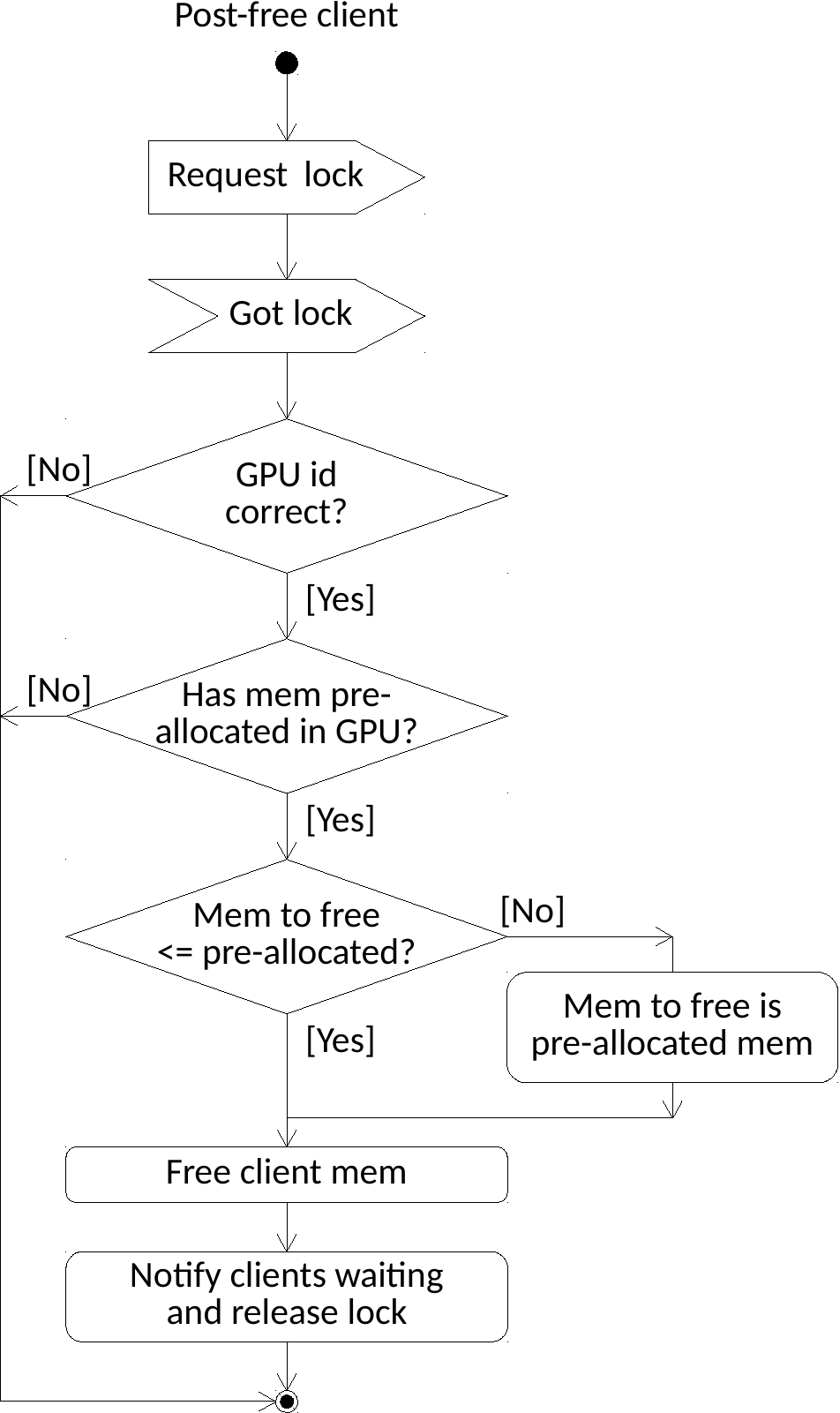}
\caption{\textcolor{black}{Post-free of memory by a schedGPU client}}
\label{fig:shared_memory_activity_postfree}
\end{figure}

\textit{Post-free}: %\label{postfree}
As shown in Figure \ref{fig:shared_memory_activity_postfree} 
for post-freeing memory using the \texttt{postCudaFree()} function, the client requests the ownership of the file lock. When the ownership of the lock is obtained the client checks that (i) the GPU requested is a valid device and (ii) the requested memory for freeing is already pre-allocated on the GPU.
If memory is freed, then the clients that may be waiting for memory are notified (refer to Section \ref{clientnotification}). 

\begin{figure}
\centering
\includegraphics[width=0.2\textwidth]{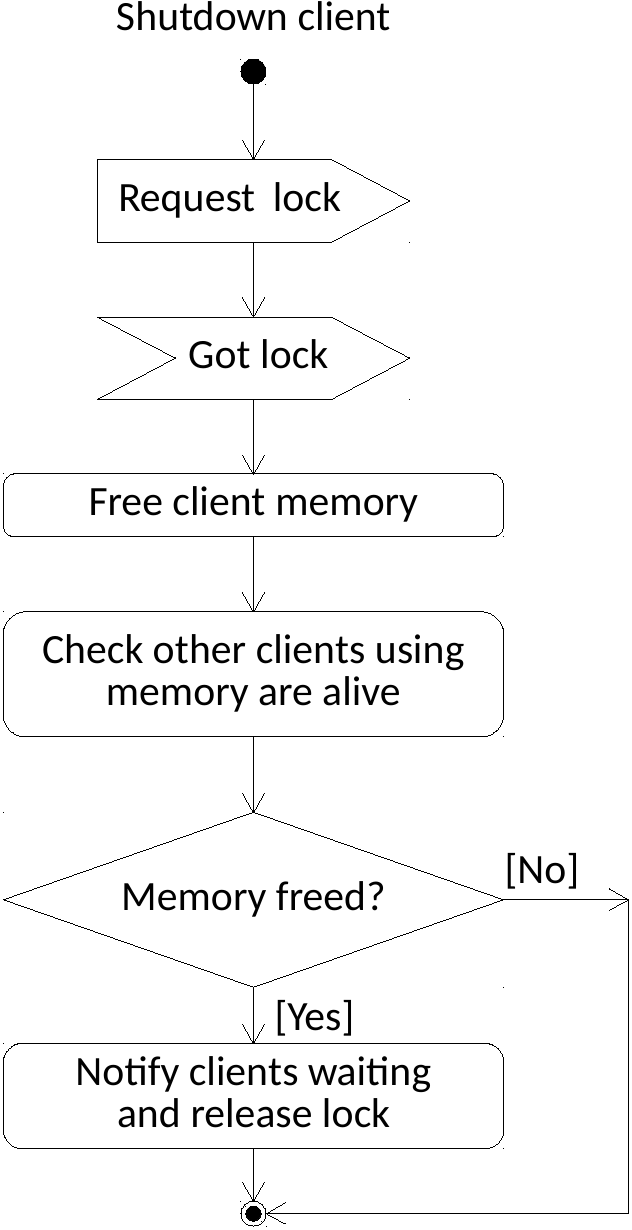}
\caption{\textcolor{black}{Shutdown of a schedGPU client}}
\vspace{-0.5cm}
\label{fig:shared_memory_activity_shutdown}
\end{figure}

%\subsection{Shutdown}
\textit{Shutdown}: As shown in Figure \ref{fig:shared_memory_activity_shutdown} the client requests the ownership of the file lock. When the ownership is obtained the client (i) ensures that it has post-freed all pre-allocated memory, and (ii) performs a sanity check whether other clients with pre-allocated memory are still alive. If memory is freed, then the waiting clients are notified (refer to Section \ref{clientnotification}). The shutdown is implicitly handled by the \texttt{postCudaFree()} function.

%The \texttt{postCudaFree()} function does not free GPU memory, instead notifies the framework that memory has been freed on the device. If \texttt{device} is less than 0, then memory pre-allocated by the framework to the application on any device is freed. If a \texttt{device} is specified, then memory in \texttt{bytes} is freed on the device. If \texttt{bytes} is set to 0, then all memory available on the selected device will be freed. 

%\begin{lstlisting}
%cudaError_t postCudaFree(int device, size_t bytes);
%\end{lstlisting}

%Error handling is implemented using the following standard CUDA error codes:
%\begin{itemize}[leftmargin=0.3cm]
% \item \texttt{cudaSuccess}: the framework has been successfully notified
% \item \texttt{cudaErrorUnknown}: indicates an internal error
%\end{itemize}

%% file: policies.tex
\begin{figure*}
\begin{center}
\subfloat[ARA]
{
	\label{fig:bv6_cuda_cpu_gpu_usage}
	\includegraphics[width=0.35\textwidth]
	{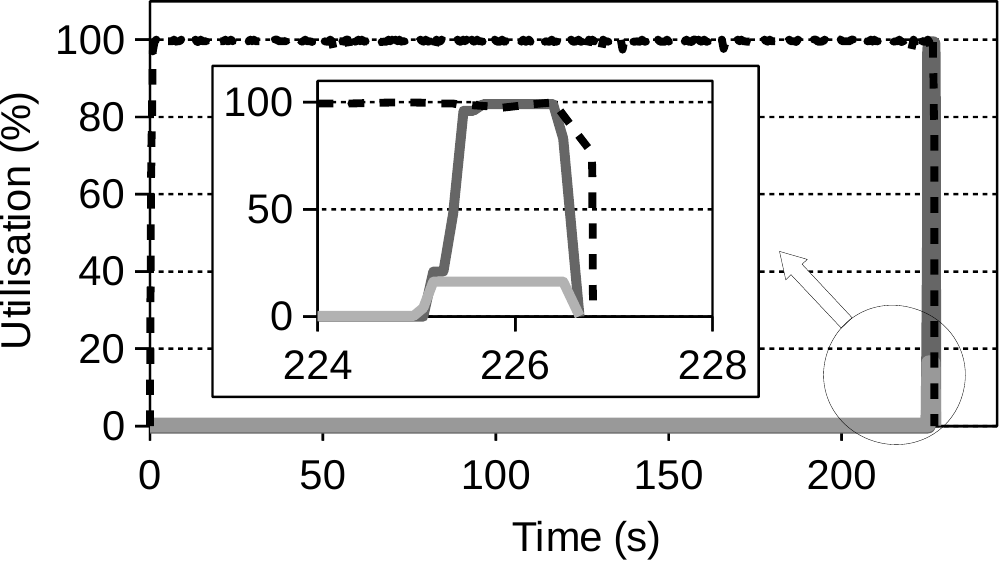}
}
\hspace{-0.3cm}
\subfloat[MUMmerGPU]
{
	\label{fig:mummergpu_cuda_cpu_gpu_usage}
	\includegraphics[width=0.302\textwidth]
	{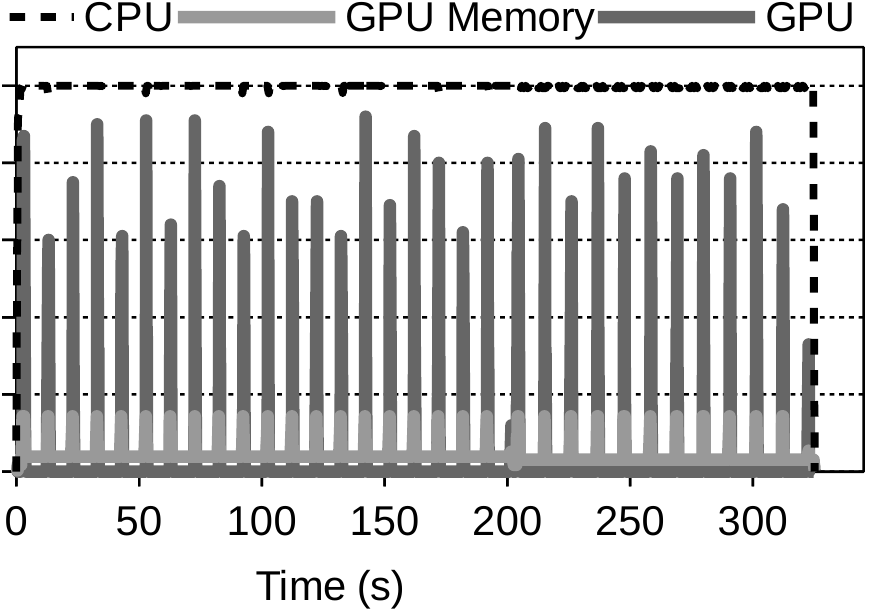}
}
\hspace{-0.3cm}
\subfloat[GPU-BLAST]
{
	\label{fig:gpublast_cuda_cpu_gpu_usage}
	\includegraphics[width=0.31\textwidth]
	{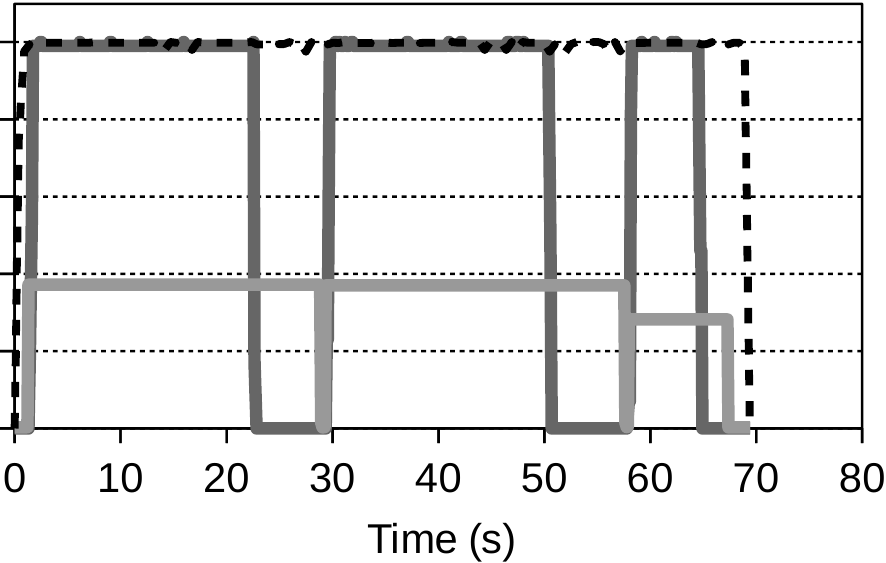}
}
\end{center}
\caption{CPU and GPU utilisation and GPU memory used for one execution of the applications.}
\vspace{-0.5cm}
\label{fig:3apps_cuda_cpu_gpu_usage}
\end{figure*}

Client applications that wait in a queue for GPU memory are notified when memory is available because another application released it. Policies are required to schedule memory requests of waiting clients.
Scheduling policies are reported for managing CPU resources~\cite{schedpolicy-3,schedpolicy-4}. Popular policies include First-In, First-Out (FIFO) and those that maximise resource utilisation. Priority-based policies are implemented to prioritise execution by the operating system. We adapted these policies in the context of GPU co-scheduling.

The FIFO policy is simple, but is limited in that if the first waiting application's GPU memory request cannot be furnished, then even if there was a subsequent waiting client that could be scheduled to access the GPU has to wait. This potentially reduces the utilisation rate of the GPU. This can be mitigated by using policies based on consumable resources. In the context of GPU co-scheduling, a policy to maximise the usage of memory on the device is ideal, which in turn increases utilisation. 

The basic version of both FIFO and Maximum Memory Utilisation (MMU) policies do not consider the quality of service offered to the clients. This requires priority of waiting applications to be accounted for to provide preferential services to applications with higher priorities. In this paper, we considered four policies, two basic policies and two priority-based policies. 

GPU utilisation can be used as a good priority criterion in policies as in CPU scheduling. However, we have chosen GPU memory utilisation since it is a more stable indicator given that it can be quantified more reliably than GPU utilisation. This is because current techniques employed on GPUs do not provide accurate estimates of utilisation as on the CPU. Therefore, the policies considered in this paper are based on GPU memory.

%\subsubsection*{Policy 1 - First-In, First-Out (FIFO)}
\textit{Policy 1 - First-In, First-Out (FIFO)}:
\textcolor{black}{Consider there are $n$ waiting clients, represented as $C = \{C_1, C_2, \cdots, C_n\}$, and that $\Delta_i$ is maximum time client $C_i$ waits for free memory}.
\textcolor{black}{In this policy, the first waiting client in the queue $C_1$ is benefited. $C_1$ waits until there is sufficient free memory.}
If memory is available, then it is preallocated to $C_1$, and if there is more free memory then the next client is served. If there is insufficient memory, then \textcolor{black}{client $C_1$} waits for a maximum time of $\Delta_1$ as it is blocked by the scheduler \textcolor{black}{(also applied for the next 3 policies).}
% %This is represented as Equation~\ref{eq:fifo}.
% %"\quad" inserts a space
% \begin{multline} \label{eq:fifo}
% preallocate(C_i) = \\
% \begin{cases}
%       \big((M_d = M_d - m_d^{C_i}) \wedge (C - \{C_i\}) \wedge \\ \quad (preallocate(C_{i+1})) \big) , \text{ if } \big((m_d^{C_i} \leq M_d) \wedge (i < n)\big)\\
%       \big((M_d = M_d - m_d^{C_i}) \wedge (C - \{C_i\}) \wedge \\ \quad exit()\big), \text{ if } \big((m_d^{C_i} \leq M_d) \wedge (i \geq n)\big)\\
%       wait(\Delta_i), \text{ otherwise}
%     \end{cases}
% \end{multline}

%\subsubsection*{Policy 2 - Maximum Memory Utilisation (MMU)}
\textit{Policy 2 - Maximum Memory Utilisation (MMU)}: The aim of this policy is to use maximum GPU memory and hence the request of the first client in the queue that can be pre-allocated memory is furnished. If no clients can be serviced, then the clients continue to wait until a subsequent client terminates and memory is available.
% \begin{multline} \label{eq:mmu}
% preallocate(C_i) = \\
% \begin{cases}
%       M_d = M_d - m_d^{C_i} \wedge C - \{C_i\} \wedge preallocate(C_{i+1}), \\ \quad \text{ if } m_d^{C_i} \leq M_d\\
%       wait(\Delta_i) \wedge preallocate(C_{i+1}) \mbox{ concurrently}, \text{ if } i < n\\
%       exit(), \text{ otherwise }
%     \end{cases}
% \end{multline}

%\subsubsection*{Policy 3 - Priority FIFO}
\textit{Policy 3 - Priority FIFO}: This policy is similar to the FIFO policy, but has a priority associated with each client. Therefore, in the queue, the clients with the highest priority are pre-allocated memory. The first client with the highest priority will be served, but if there is insufficient memory to serve this request or there is more memory available after serving a request, then a following client with the same priority is served.
% \begin{multline} \label{eq:priority_fifo}
% preallocate(C_i^p) = \\
% \begin{cases}
%       \big((M_d = M_d - m_d^{C_i^p}) \wedge (C - \{C_i^p\}) \wedge \\ \quad (preallocate(C_{i+1}^p)) \big) , \text{ if } \big((m_d^{C_i^p} \leq M_d) \wedge (i < n)\big)\\
%       \big((M_d = M_d - m_d^{C_i^p}) \wedge (C - \{C_i^p\}) \wedge \\ \quad exit()\big), \text{ if } \big((m_d^{C_i^p} \leq M_d) \wedge (i \geq n)\big)\\
%       wait(\Delta_i), \text{ otherwise}
%     \end{cases}
% \end{multline}

%\subsubsection*{Policy 4 - Priority MMU}
\textit{Policy 4 - Priority MMU}: This policy is similar to the MMU policy, but has a priority associated with each client. To maximise GPU memory usage the request of the first client with the highest priority in the queue that can be pre-allocated memory is furnished. If no clients in the queue with the highest priority can be attended to, then clients continue waiting until a subsequent client terminates and more memory is available. %This is shown in Equation~\ref{eq:priority_mmu}.
% \begin{multline} \label{eq:priority_mmu}
% preallocate(C_i^p) = \\
% \begin{cases}
%       M_d = M_d - m_d^{C_i^p} \wedge C - \{C_i^p\} \wedge preallocate(C_{i+1}^p),  \\ \quad \text{ if } m_d^{C_i^p} \leq M_d\\
%       wait(\Delta_i) \wedge preallocate(C_{i+1}^p) \mbox{ concurrently}, \text{ if } i < n\\
%       exit(), \text{ otherwise }
%     \end{cases}
% \end{multline}

The implications of using the above policies for executing workloads is explored in the subsequent sections.

%% file: setup.tex
The hardware platform and the {\color{black} benchmarks and }use-cases employed for validating the feasibility of schedGPU are presented in this section. 

%\subsection{Hardware Platform}
\textit{Hardware Platform:} 
The experimental test-bed used for our experiments is one 1027GR-TRF Supermicro server comprising two Intel Xeon hexa-core processors E5-2620 v2 (Ivy Bridge) operating at 2.1~GHz and 32~GB of DDR3 SDRAM memory at 1,600 MHz. One NVIDIA Tesla K20m GPU which has 4,799~MiB of memory is available on the server. The CentOS 6.4 operating system and the CUDA 7.5 with NVIDIA driver 352.39 is used.

\textcolor{black}{
\textit{Benchmarks:}
We evaluate the performance of schedGPU using two popular GPU benchmark suites, namely Rodinia~\cite{Rodinia} and Parboil~\cite{Parboil}}.

%\subsection{Use-cases}
%\label{use-cases}
\textit{Use-cases:}
Three real-world applications are considered as use-cases in this paper. The first is a catastrophe risk simulation employed in the financial risk industry, referred to as Aggregate Risk Analysis (ARA) \cite{ARA-1}. This simulation computes a key risk metric, namely Probable Maximum Loss (PML) on an industry size input comprising 150,000 catastrophic event trials and a collection of one thousand events and their corresponding losses.

The second and third are applications for aligning DNA sequences in bioinformatics. The second application, which is referred to as MUMmerGPU%\footnote{\url{http://www.mummergpu.sourceforge.net}}
~\cite{mummergpu}, is used for aligning DNA sequence data to a reference sequence which is useful in genotyping and genomics. In our experiments, the search pattern is a sequence length of 25 base pairs that is matched against the reference, which is a complete genome of Bacillus Anthracis allowing up to five differences in an alignment for 500,000 reads. 

The third application is referred to as the GPU Basic Local Alignment Search Tool (GPU-BLAST)%\footnote{\url{http://archimedes.cheme.cmu.edu/biosoftware.html}}
~\cite{gpublast}. The application searches a database of proteins for a nucleotide with a sequence length of 5,000.

The use-cases were chosen based on the following three observations from Figure \ref{fig:3apps_cuda_cpu_gpu_usage}, which shows the CPU and GPU utilisation and the GPU memory in use during execution. Firstly, \textit{low GPU utilisation}. ARA, in Figure~\ref{fig:bv6_cuda_cpu_gpu_usage}, uses GPU acceleration for a short time period at the end of the simulation. For the given input, over 16\% of GPU memory is used and therefore, up to a maximum of 6 concurrent instances of the application can be safely executed on this GPU without potential GPU memory allocation errors (in this paper, we refer to this as `maximum concurrent instances'). Such concurrent applications that have low GPU utilisation are ideal candidates for schedGPU since the framework can coordinate the access of multiple applications to the GPU, which otherwise would execute sequentially. 

Secondly, \textit{moderate GPU utilisation}. MUMmerGPU, in Figure~\ref{fig:mummergpu_cuda_cpu_gpu_usage}, harnesses GPU acceleration at regular intervals. For the given input, the GPU is used for approximately 50\% of the total execution time and the maximum GPU memory used is nearly 15\% allowing for up to 6 parallel instances of the application to be reliably executed. Concurrent executions of moderate GPU utilising applications are again ideal candidates for schedGPU since the framework can maximise the number of these applications safely using the GPU.

Thirdly, \textit{high GPU utilisation (and GPU memory is still available)}. GPU-BLAST (Figure~\ref{fig:gpublast_cuda_cpu_gpu_usage}) uses the GPU nearly 80\% of the total execution time, but for the given input maximum GPU memory used is over 36\% of total available memory. This allows for safe execution of up to 2 concurrent application instances. This is not an ideal candidate for schedGPU, however performance gains may be obtained when GPU memory usage drops below 30\% towards the end of the execution.

%% file: evaluation.tex
In this section, we present the experiments carried out for validating the feasibility of schedGPU. For this we (i) evaluate the overheads associated with the client-server and shared memory approaches, \textcolor{black}{(ii) analyse the performance using popular GPU benchmark suites,} (iii) highlight the benefits of employing schedGPU to improve the throughput of concurrent executions of an individual application, and (iv) consider the performance gain of workloads comprising multiple applications. 

\subsection{Overhead of the approaches}
\begin{figure}
\centering
\includegraphics[width=0.45\textwidth]{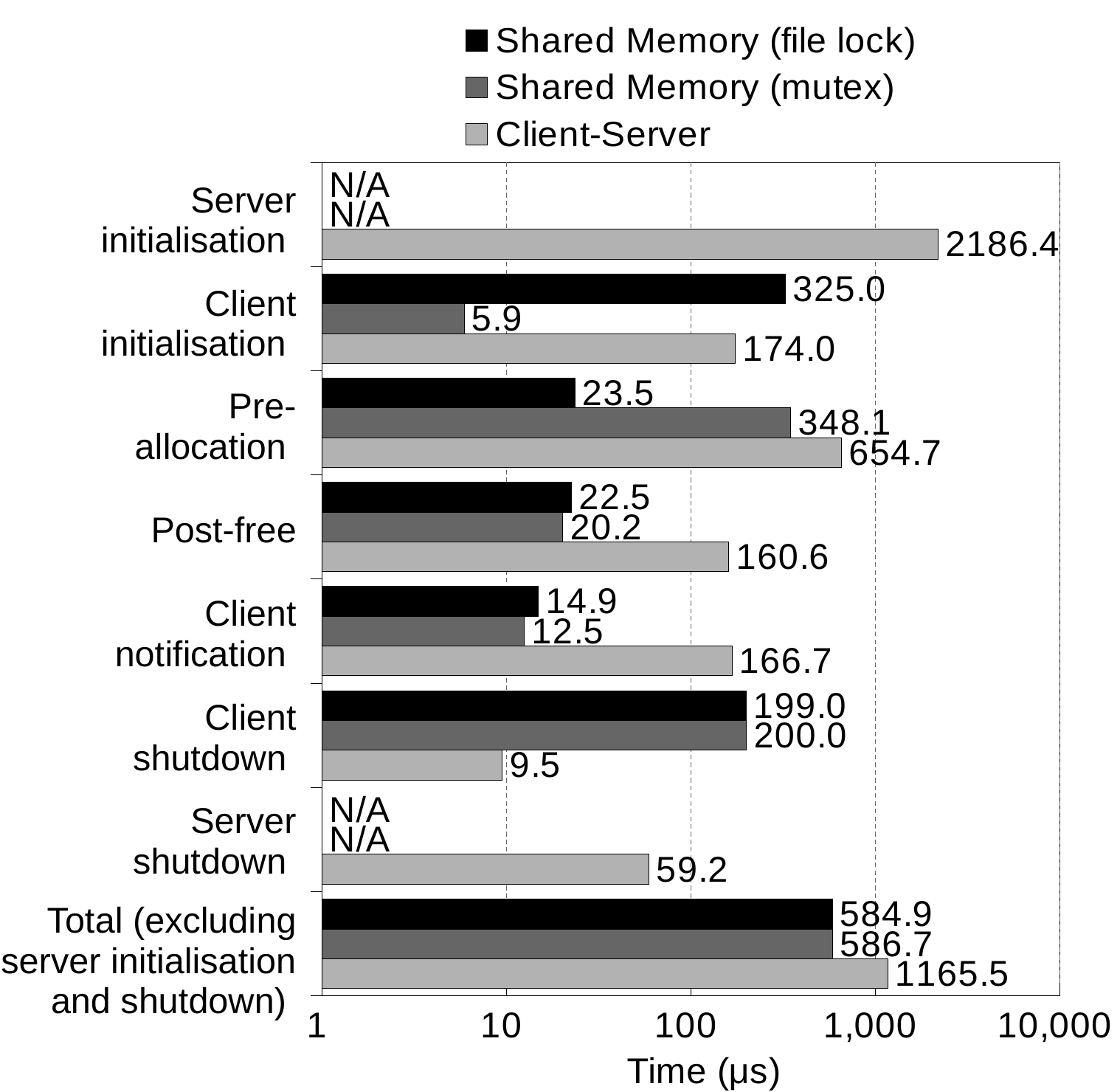}
\caption{Comparison of the stages of the schedGPU life cycle for the client-server and shared memory approaches.}
\vspace{-0.5cm}
\label{fig:frameworks_comparison}
\end{figure}

Figure~\ref{fig:frameworks_comparison} compares different stages of the schedGPU life cycle. Both the client-server and shared memory approaches are considered. For the former, an additional server initialisation and server shutdown stages are required since these are distributed between the client and the server. For the latter, initialisation and shutdown are carried out by the client since no servers are present.  

It is observed that the server initialisation and server shutdown stages for the client-server approach are costly in terms of time although they occur only once. The client initialisation and client shutdown stages are however shorter. Regardless, even when excluding the time for initialising and shutting down the server, the total time taken by the client-server is nearly twice as taken by the shared memory approach. This is because communications over \textcolor{black}{UNIX} sockets introduce an overhead. Therefore, only the shared memory approach is employed in the experiments considered in subsequent sections.  

Both shared memory approaches using mutexes and file locks offer similar performance. In the initialisation stage, the file locks method requires more time since the notification protocol using system signals needs to be set-up. With mutexes the notification protocol using conditions is set-up in the pre-allocation stage and hence an increase in time for the pre-allocation stage is noted. 

Given that both the shared memory approaches have similar performance, in the following sections we consider the file lock method since it is more robust than mutexes by avoiding the problem of abandonment. %considered in Section~\ref{synchronisingaccess}. 

\textcolor{black}{The overhead of the different implementations is less than one millisecond. This does not have any impact on long running applications. However, the shared memory implementation does not require an additional service to run on the server. This is valued by administrators of production systems to keep the number of services running on a server to a minimum for security reasons.}

\begin{figure*}
\centering
\includegraphics[width=0.9\textwidth]{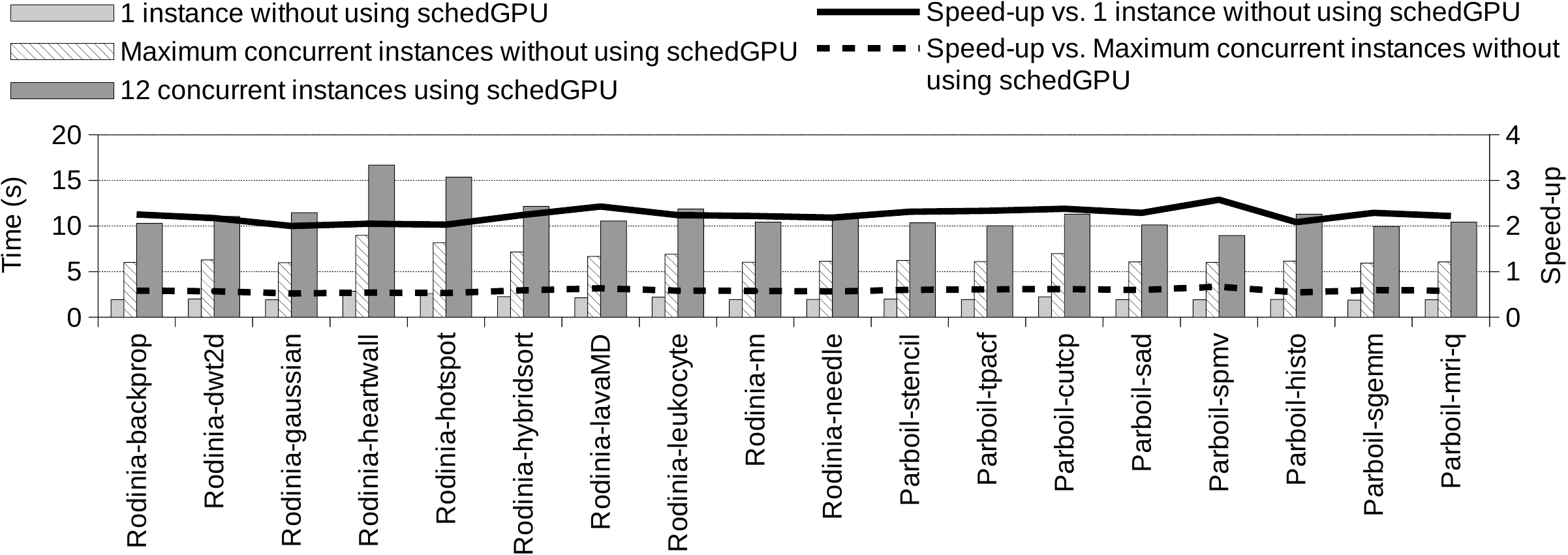}
\caption{\textcolor{black}{Speed-up using schedGPU when concurrently running 12 instances of benchmarks selected from Rodinia and Parboil suites.}}
\label{fig:rodinia_parboil}
\vspace{-0.45cm}
\end{figure*}

\textcolor{black}{\subsection{schedGPU with Benchmark Suites}\label{benchmark_suites}}

\textcolor{black}{Figure~\ref{fig:rodinia_parboil} shows the execution time and speed-up of 10 Rodinia and 8 Parboil benchmarks when schedGPU is employed. The tests were run in three different scenarios. The first scenario does not use schedGPU and one instance of the benchmark is safely executed at a time. Hence, if 12 instances of the benchmark were required to be executed, then they are executed sequentially. The second scenario, similarly does not use schedGPU. However, we manually packed tasks to maximise GPU memory usage. This is not realistic, but was pursued for the sake of comparison, as workload schedulers do not know GPU memory required by an application in advance. The third scenario employs schedGPU and safely runs multiple instances of the benchmark.}

\textcolor{black}{Multiple instances of an application or multiple applications are executed concurrently using schedGPU. Therefore, the overall execution time of all instances/applications is reduced when compared to running them sequentially, which is referred to as performance speed-up. However, the execution time of an individual instance is not improved and optimising performance of individual applications is not within scope of this paper.}

\textcolor{black}{We observe that using schedGPU there is over a 2x speed-up when compared against the first scenario that does not employ schedGPU. On the contrary, compared to the second scenario, the execution time is increased 40\% on average when using schedGPU. These benchmarks have short execution times (2 seconds on average), therefore, it is more difficult to compensate the overhead introduced by schedGPU with the potential gain of concurrently using the GPU. However, this is because we have manually packed tasks to maximise the usage of GPU memory, but as previously noted this is not realistic. In the next section, we use production codes instead of benchmarks to demonstrate the benefits of schedGPU.\\}

\subsection{schedGPU Performance Gain on Use-cases}
We further explore performance in terms of utilisation of GPU resources, speed-up and throughput in the following two ways: (i) on an experimental environment, to study mechanisms to achieve maximum performance of the three use-cases with and without schedGPU, and (ii) on a production environment, to assess the potential of schedGPU using real-world workloads.

\subsubsection{Concurrent Execution of Individual Applications}
Three schedGPU functions considered in Section~\ref{apifunctions} were included in the three applications. The initialisation function was included at the beginning of the CUDA program, the pre-allocation function was inserted before the CUDA memory allocations and the post-free function was placed after CUDA release memory calls. %\textcolor{red}{\sout{This requires minimum programming efforts and validates the simplicity of our approach.}} 
Up to 12 instances of each application were concurrently executed (the number of CPU cores in the experimental test-bed). 

\begin{figure*}
\centering
\includegraphics[width=0.92\textwidth]{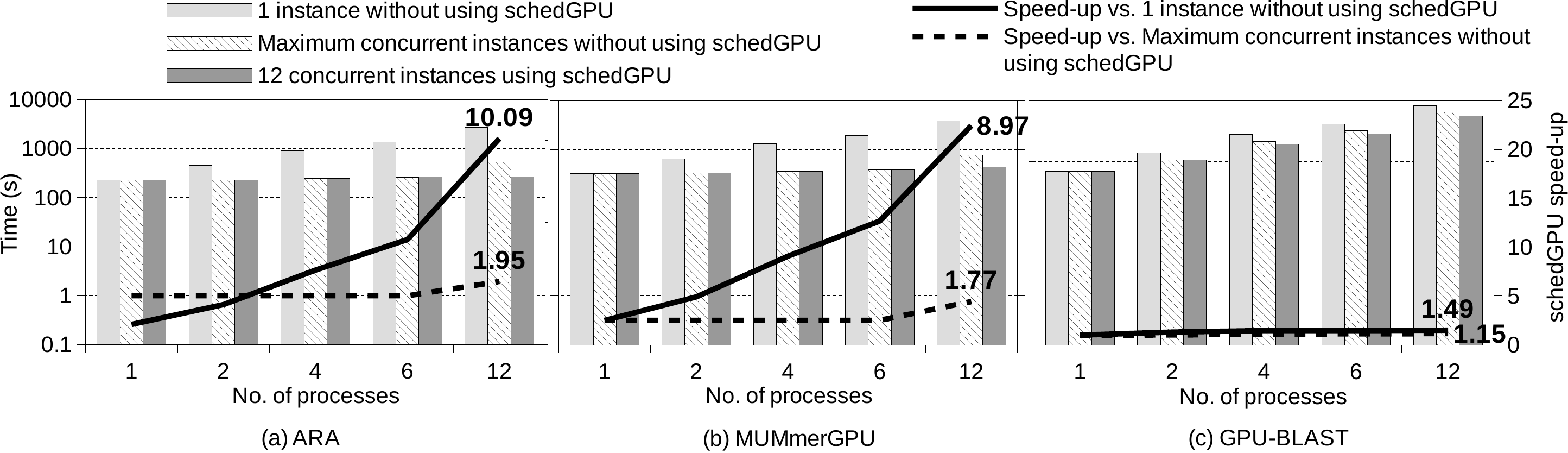}
\caption{Speed-up using schedGPU when varying number of instances of an application are concurrently executed.}
\vspace{-0.25cm}
\label{fig:3apps_cuda_vs_schedgpu_throughput}
\end{figure*}

Figure~\ref{fig:3apps_cuda_vs_schedgpu_throughput} shows the improvement in execution time and speed-up of the three applications when schedGPU is employed.
\textcolor{black}{The three scenarios of Section~\ref{benchmark_suites} were considered again}.
%carregon-v3\textcolor{red}{\sout{Similar to Section~\ref{benchmark_suites}, three scenarios were considered. The first scenario does not use schedGPU and only one instance of the application can be safely executed at a time. Hence, if 12 instances of an application were required to be executed, then they are executed sequentially. The second scenario, similarly does not use schedGPU. However, we have manually packed tasks to maximise the usage of GPU memory. This is done for the purpose of comparison and is not realistic as the workload managers do not know in advance the GPU memory required by the application. The third scenario employs schedGPU and can safely run multiple instances of the application.}}   
It is immediately inferred that when comparing our proposed approach (the third scenario) using schedGPU against (i) the first scenario that does not employ schedGPU there is a 10x speed-up for ARA, nearly 9x speed-up for MUMmerGPU and close to 1.5x speed-up for GPU-BLAST, and (ii) the second scenario the speed-up is approximately doubled when running 12 concurrent instances of ARA and MUMmerGPU. These applications have low and moderate GPU utilisation allowing schedGPU to take advantage of the time periods that the GPU remains under-utilised. During these time periods schedGPU services instances that request the GPU to maximise GPU utilisation.
In the second scenario, the execution of large number of instances of an application (more than 6 for ARA and MUMmerGPU respectively) at the same time will not be possible due to insufficient GPU memory. schedGPU still outperforms this unrealistic scenario of manually packing tasks. Not only is it feasible to execute large number of instances using schedGPU, but also a profitable speed-up is noted. 

However, there is only a small improvement in performance for GPU-BLAST with the execution of 4 concurrent instances achieving maximum speed-up. The application has high GPU memory utilisation almost during all of its execution (over 30\% on average as shown in Figure~\ref{fig:gpublast_cuda_cpu_gpu_usage}). Therefore, there is insufficient GPU memory for boosting performance of concurrent instances. Nonetheless, schedGPU yields a small improvement in performance by making use of any spare GPU memory. 

%\begin{figure*}
%\centering

%\subfloat[ARA]
%{
%	\label{fig:bv6_cuda_vs_schedgpu_throughput}
%	\includegraphics[width=0.345\textwidth]
%	{figures/bv6_cuda_vs_schedgpu_throughput.pdf}
%} 
%\hfill
%\subfloat[MUMmerGPU]
%{
%	\label{fig:mummergpu_cuda_vs_schedgpu_throughput}
%	\includegraphics[width=0.345\textwidth]
%	{figures/mummergpu_cuda_vs_schedgpu_throughput.pdf}
%}
%\hfill
%\subfloat[GPU-BLAST]
%{
%	\label{fig:gpublast_cuda_vs_schedgpu_throughput}
%	\includegraphics[width=0.345\textwidth]
%	{figures/gpublast_cuda_vs_schedgpu_throughput.pdf}
%}
%\caption{Execution time and speed-up obtained using schedGPU when varying number of instances of the same application that are concurrently executed}
%\label{fig:3apps_cuda_vs_schedgpu_throughput}
%\end{figure*}

\begin{figure*}
\begin{center}
\subfloat[ARA running 12 concurrent instances]
{
	\label{fig:bv6_schedgpu_cpu_gpu_usage}
	\includegraphics[width=0.82\textwidth]
	{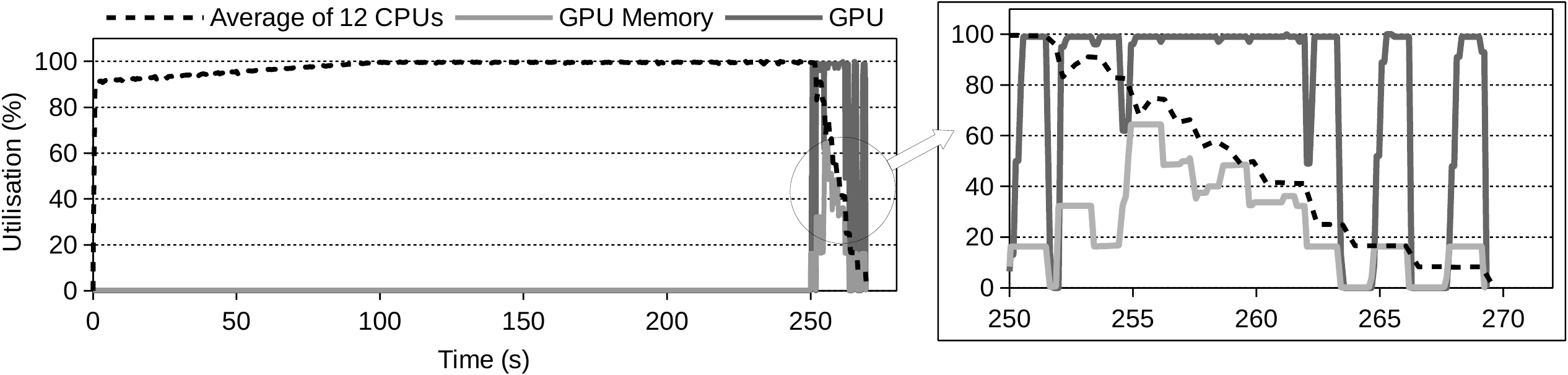}
} 
\hfill
\subfloat[MUMmerGPU running 12 concurrent instances]
{
	\label{fig:mummergpu_schedgpu_cpu_gpu_usage}
	\includegraphics[width=0.82\textwidth]
	{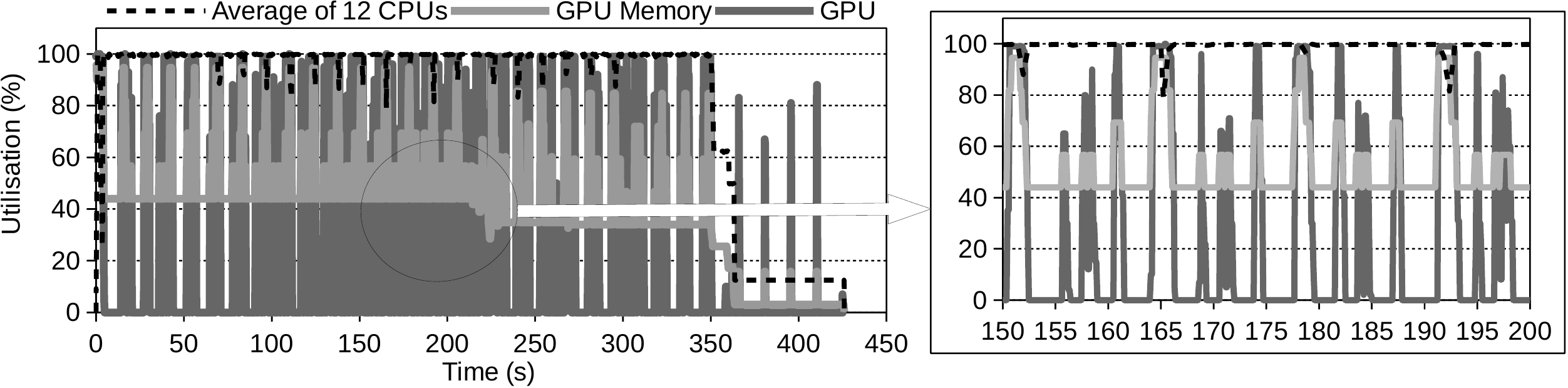}
}
\hfill
\subfloat[GPU-BLAST running 4 concurrent instances]
{
	\label{fig:gpublast_cuda_vs_schedgpu_usage}
	\includegraphics[width=0.82\textwidth]
	{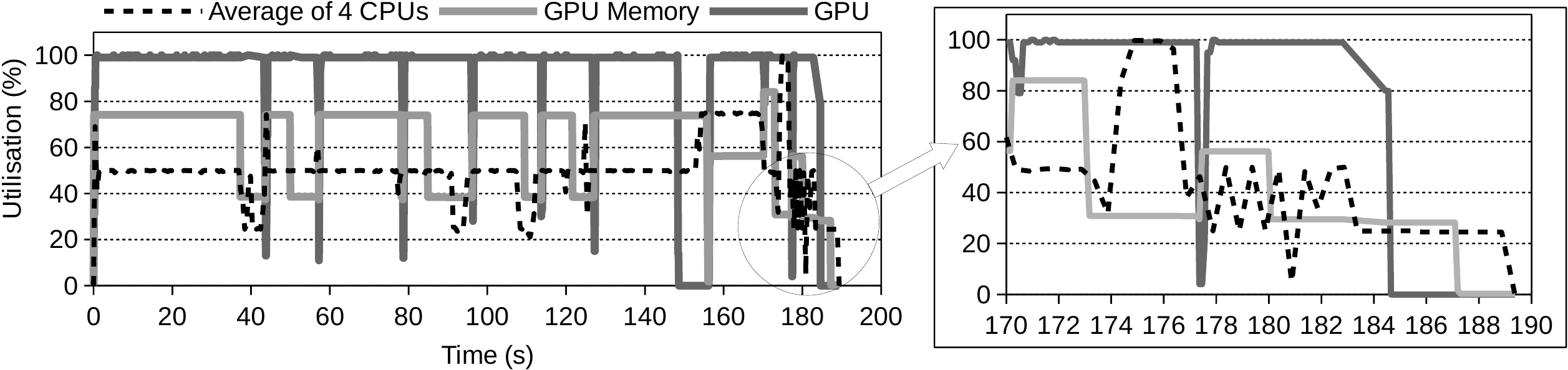}
}
\end{center}
\caption{CPU and GPU usage when running concurrent instances of the applications using schedGPU.}
\vspace{+0.25cm}
\label{fig:3apps_schedgpu_cpu_gpu_usage}
\end{figure*}

Figure~\ref{fig:3apps_schedgpu_cpu_gpu_usage} shows the average CPU, GPU and GPU memory utilisation when maximum speed-up is obtained for the three applications using schedGPU. The amount of GPU memory utilised by each application is indicated in the figures. GPU utilisation is maximised, which in turn results in an observed speed-up.   

\begin{figure*}
\centering
\includegraphics[width=0.9\textwidth]{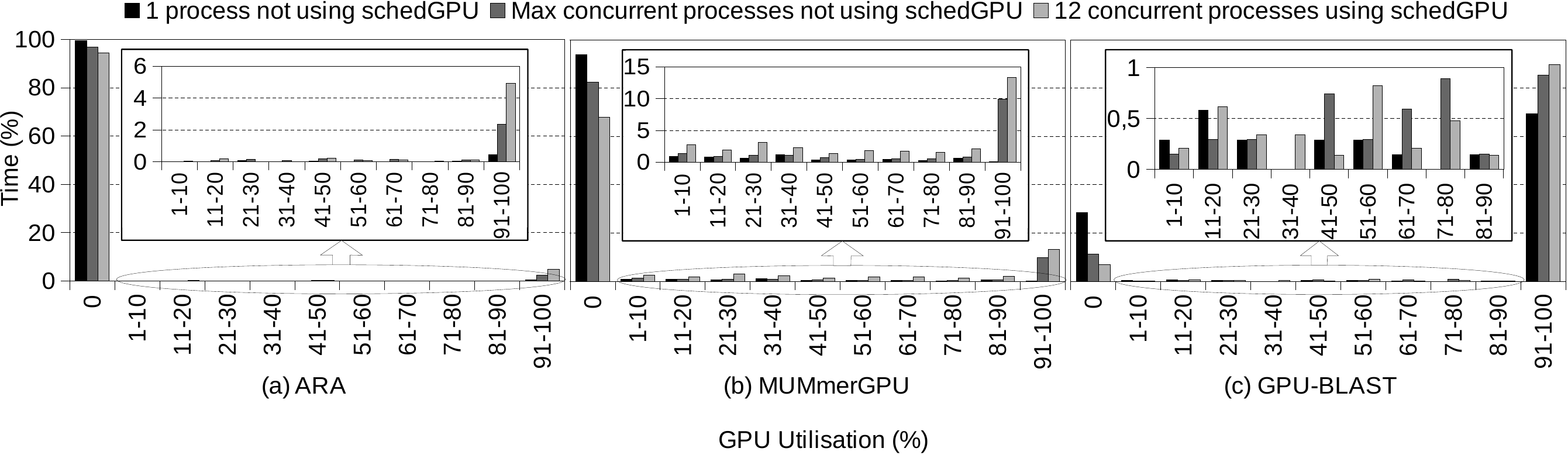}
\caption{Frequency distribution of GPU utilisation when executing the application with and without schedGPU.}
\vspace{-0.5cm}
\label{fig:3apps_cuda_vs_schedgpu_gpu_usage}
\end{figure*}

Figure \ref{fig:3apps_cuda_vs_schedgpu_gpu_usage} shows the frequency distribution of GPU utilisation for the three applications. For all applications it is observed that the amount of time the GPU achieves between 91\% and 100\% utilisation is increased (the time the GPU is not utilised decreases - 0\%) and yields a speed up as shown in Figure \ref{fig:3apps_cuda_vs_schedgpu_throughput}. This validates that schedGPU can improve the utilisation of resources.

Table \ref{Table:gpu_usage_1} shows the average GPU utilisation and GPU memory utilisation for the applications when one instance of the application is executed without using schedGPU, running the maximum number of concurrent instances of the application supported without using schedGPU and 12 concurrent instances using schedGPU are employed. It is evident that schedGPU has superior performance since GPU utilisation is improved over 10 times for a single instance and nearly 2 times over six instances for ARA. Similarly, GPU memory utilisation is improved over 10 times for a single instance of MUMmerGPU and nearly 2.25 times over six instances of ARA. The memory utilisation of GPU-BLAST is high without using schedGPU leaving little room for optimisation. However, a small improvement is noted. 

\begin{table}
\caption{Comparison of GPU utilisation and GPU memory utilisation when executing the use-cases}
\label{Table:gpu_usage_1}
\begin{center}
\begin{tabular}{ | l | p{1.8cm} | p{1.8cm} | p{1.8cm} | }
\hline
                                  	&\multicolumn{3}{|p{5.4cm}|}{Average GPU Utilisation (\%), Average GPU Memory Utilisation (\%)} \\ \cline{2-4}
 	& 				& Maximum  	&  \\ 
\multicolumn{1}{|c|}{Application}	& 1 instance without using schedGPU &	concurrent instances without using schedGPU & 12 concurrent instances using schedGPU\\

\hline
ARA									& 0.51, 0.37					& 2.74, 0.90				& 5.26, 2.02		\\ \hline
MUMmerGPU							& 2.46, 4.04					& 12.73, 26.13 				& 20.96, 41.59		\\ \hline
GPU-BLAST							& 69.48, 33.73					& 86.28, 63.71				& 90.24, 70.09		\\ \hline
\end{tabular}
\end{center}
\vspace{-0.5cm}
\end{table}

%\textcolor{red}{\sout{Multiple instances of an application or multiple applications are executed concurrently using schedGPU. Therefore, the overall execution time of all instances/applications is reduced when compared to running them sequentially, which is referred to as performance speed-up. However, the execution time of an individual instance is not improved and optimising performance of individual applications is not within scope of this paper.}}

\textcolor{black}{Table~\ref{Table:execution_time} shows that the individual performance of an application is not improved by co-scheduling. We compare the execution time of the individual application when (i) running 1 instance without using schedGPU and (ii) the average of 1 instance when running 12 concurrent instances using schedGPU. The execution time of an individual instance is not improved by co-scheduling, but there is a collective performance gain when running multiple instances concurrently.}

\begin{table}
\caption{\textcolor{black}{Comparison of execution time when (i) running 1 instance of each use-case without using schedGPU and (ii) the average of 1 instance of the use-case when running 12 concurrent instances using schedGPU}}
\label{Table:execution_time}
\begin{center}
\begin{tabular}{ | l | p{2.5cm} | p{2.5cm} |}
\hline
\multicolumn{1}{|c|}{\textcolor{black}{Application}}	& \textcolor{black}{1 instance without using schedGPU} &	\textcolor{black}{12 concurrent instances using schedGPU}\\
\hline
\textcolor{black}{ARA}		& \textcolor{black}{226.847}					& \textcolor{black}{260.108}	\\ \hline
\textcolor{black}{MUMmerGPU}	& \textcolor{black}{322.799}					& \textcolor{black}{370.169}	\\ \hline
\textcolor{black}{GPU-BLAST}	& \textcolor{black}{69.144}				& \textcolor{black}{184.751}	\\ \hline
\end{tabular}
\end{center}
\vspace{-0.5cm}
\end{table}

\subsubsection{Workloads Comprising Multiple Applications}

\begin{figure}[!ht]
\vspace{0.25cm}
\begin{center}
\subfloat[Using FIFO policy]
{
	\label{fig:workload1_cpu_gpu_usage_FIFO}
	\includegraphics[width=0.49\textwidth]
	{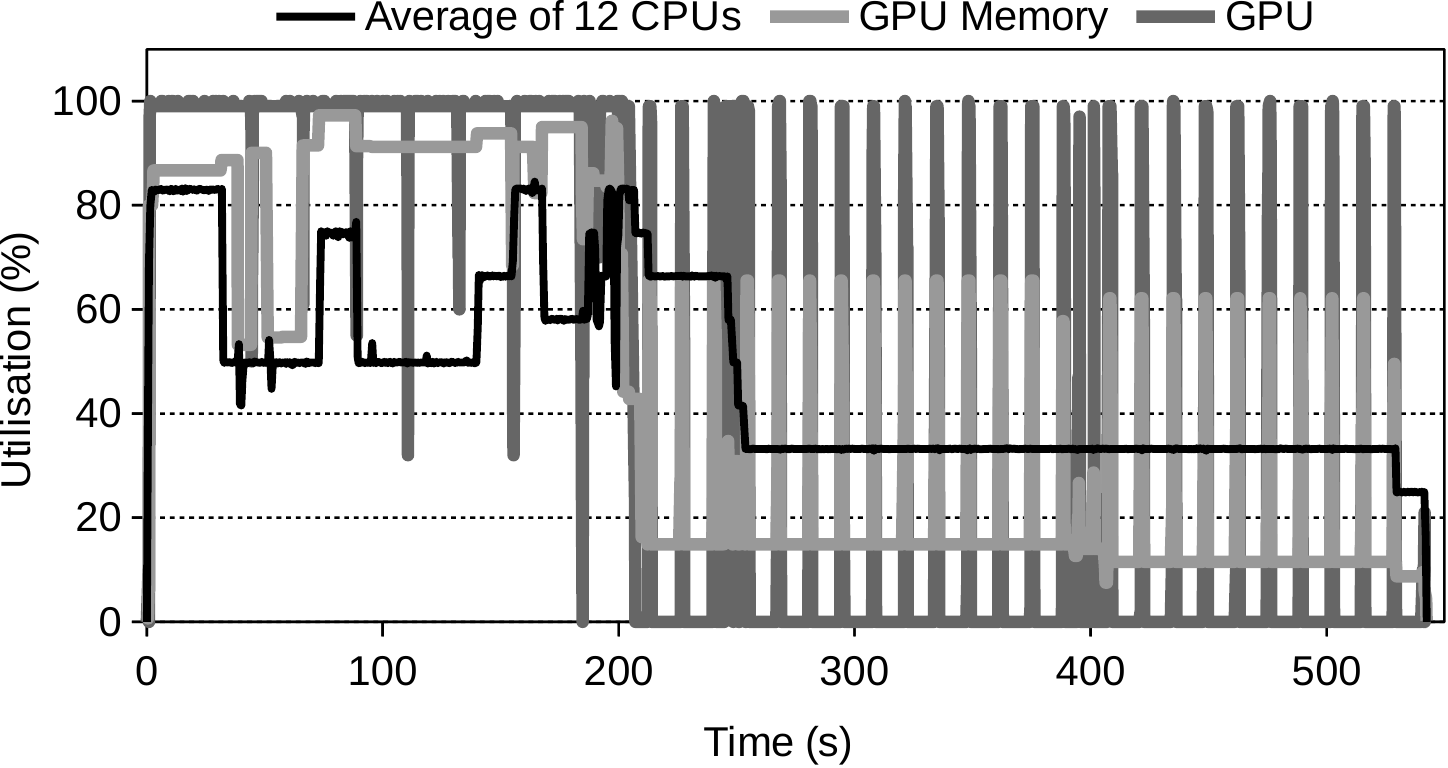}
} 
\hfill
\subfloat[Using MMU policy]
{
	\label{fig:workload1_cpu_gpu_usage_MMU}
	\includegraphics[width=0.49\textwidth]
	{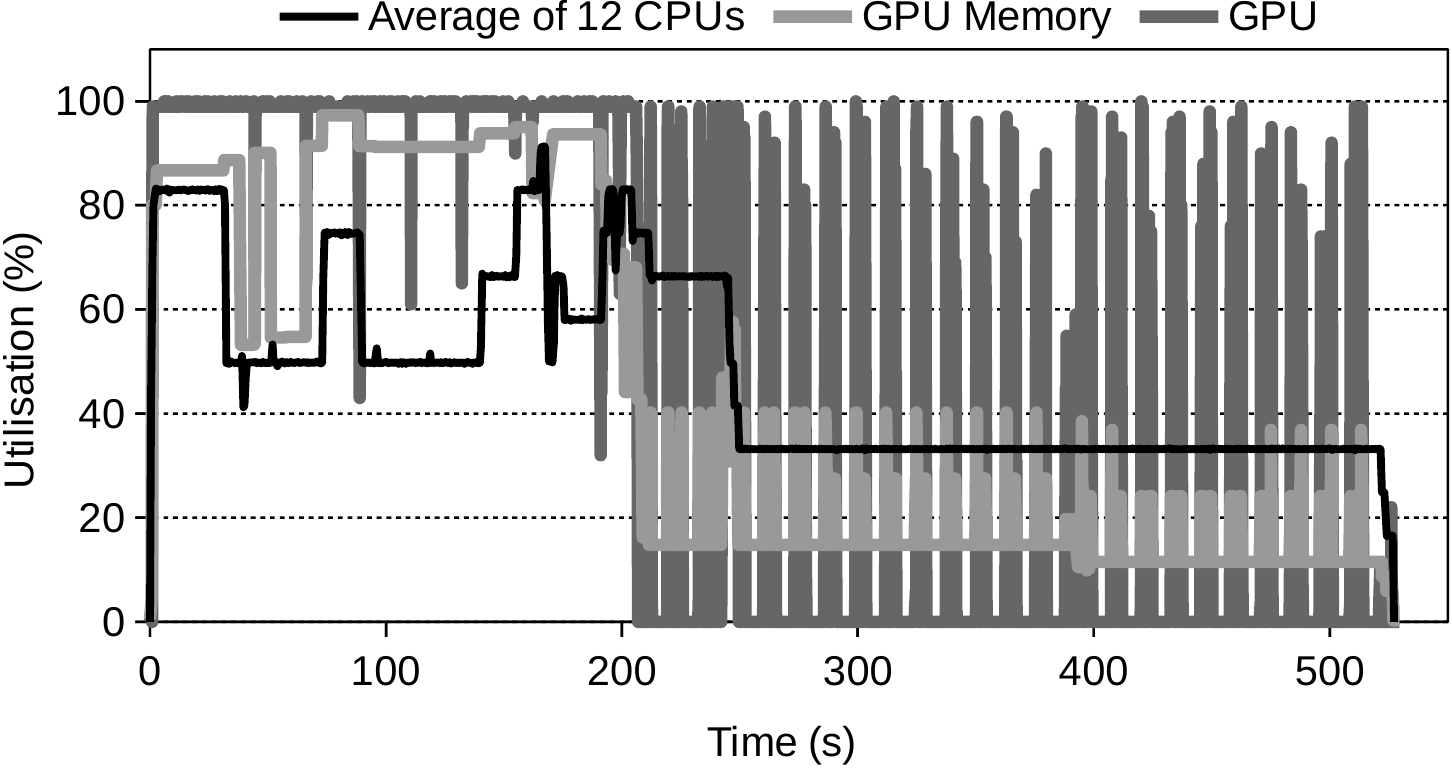}
}
\hfill
\subfloat[Using Priority FIFO policy]
{
	\label{fig:workload1_cpu_gpu_usage_PriorityFIFO}
	\includegraphics[width=0.49\textwidth]
	{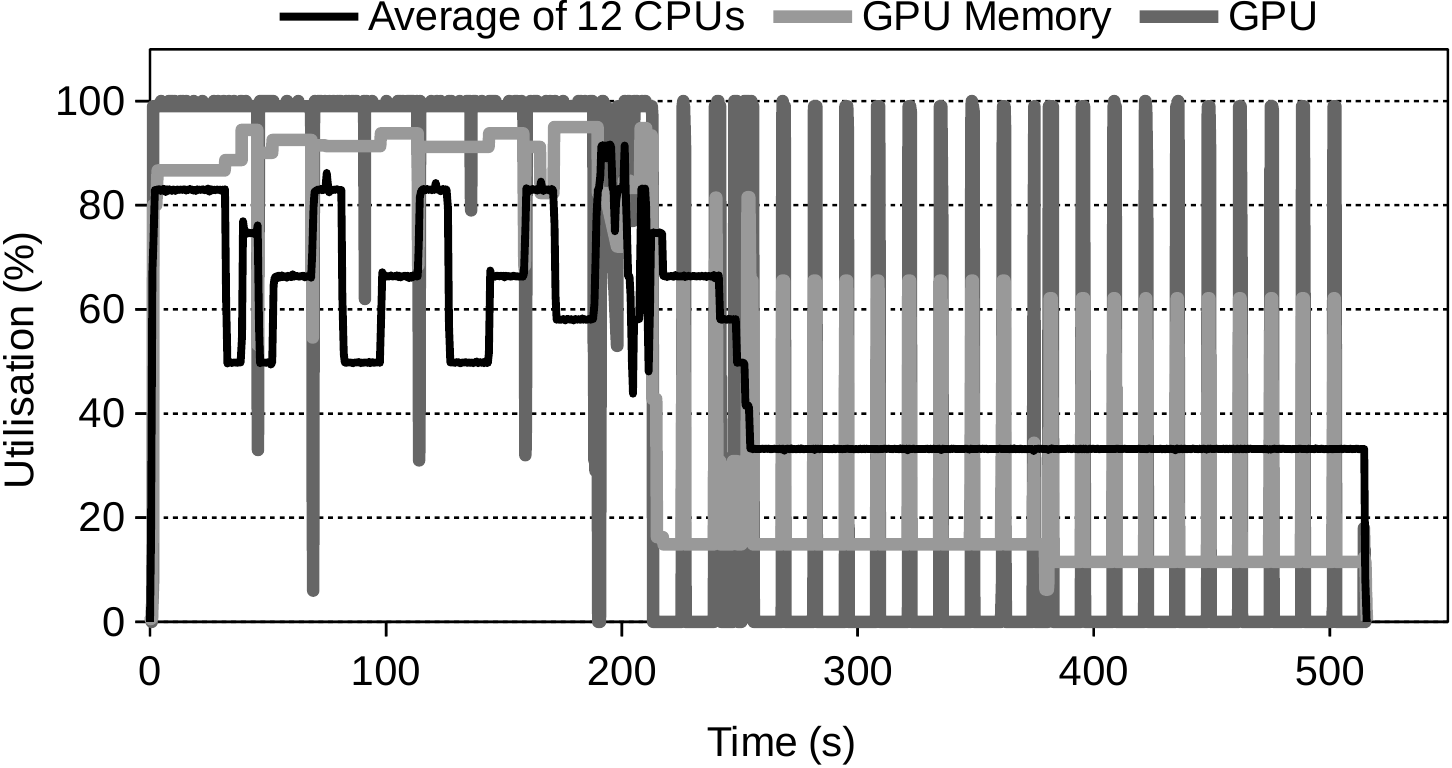}
} 
\hfill
\subfloat[Using Priority MMU policy]
{
	\label{fig:workload1_cpu_gpu_usage_PriorityMMU}
	\includegraphics[width=0.49\textwidth]
	{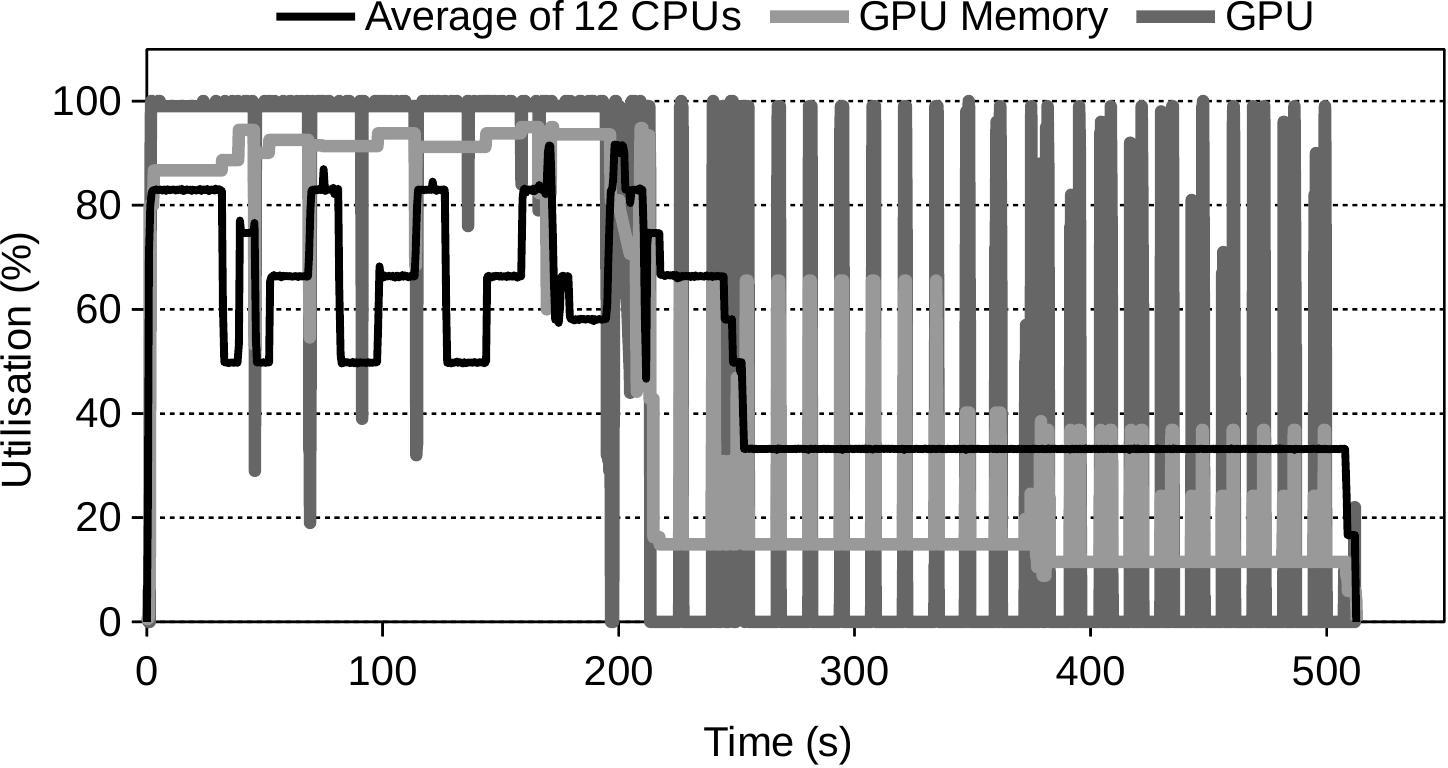}
}
\end{center}
\caption{CPU and GPU usage when running a workload using schedGPU for different client notification policies.}
\label{fig:workload1_cpu_gpu_usage}
\end{figure}

\begin{figure}[!ht]
\vspace{0.20cm}
\begin{center}
\subfloat[Using FIFO policy]
{
	\label{fig:workload1_cpu_usage_FIFO}
	\includegraphics[width=0.52\textwidth]
	{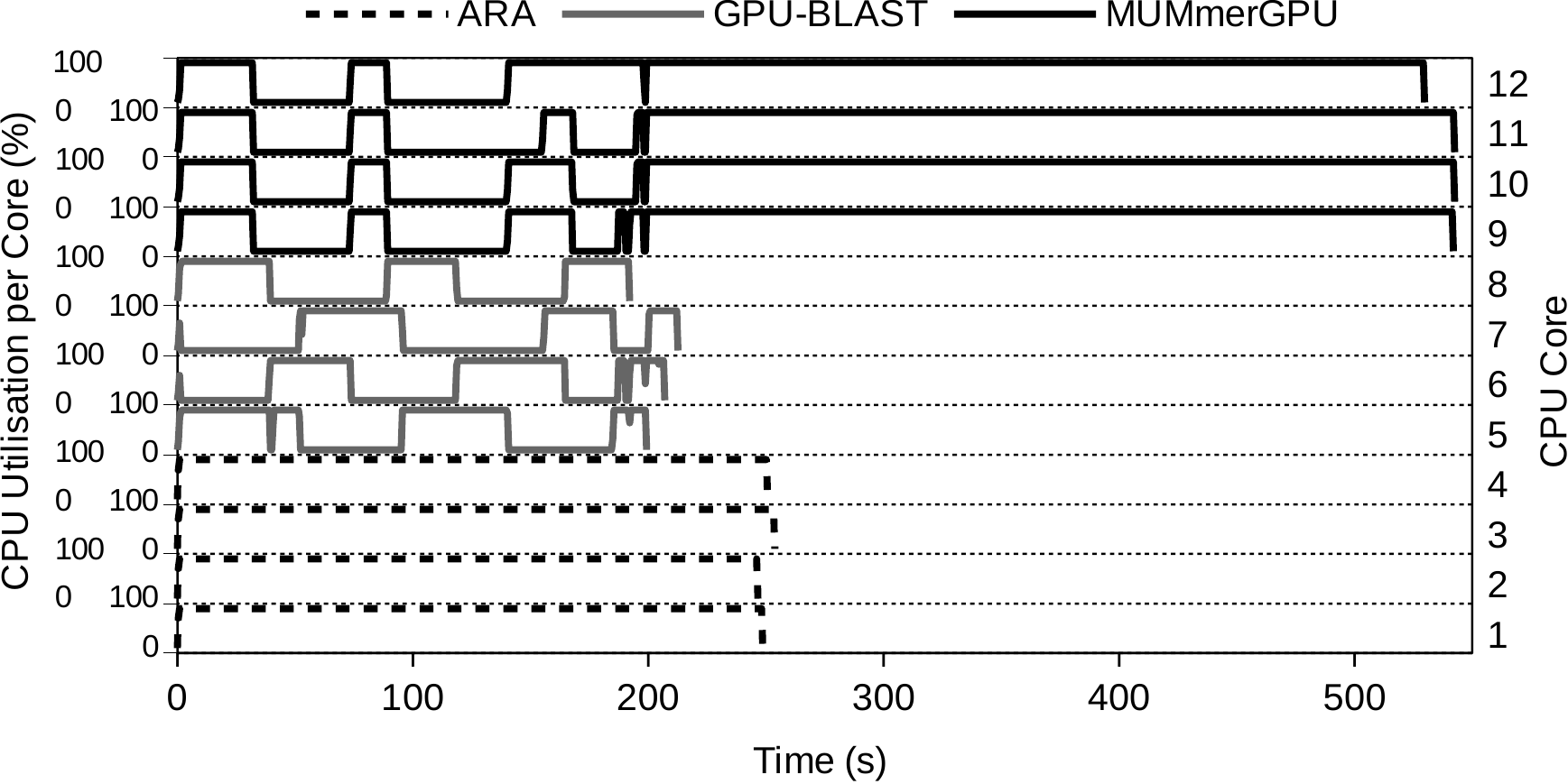}
} 
\hfill
\subfloat[Using MMU policy]
{
	\label{fig:workload1_cpu_usage_MMU}
	\includegraphics[width=0.52\textwidth]
	{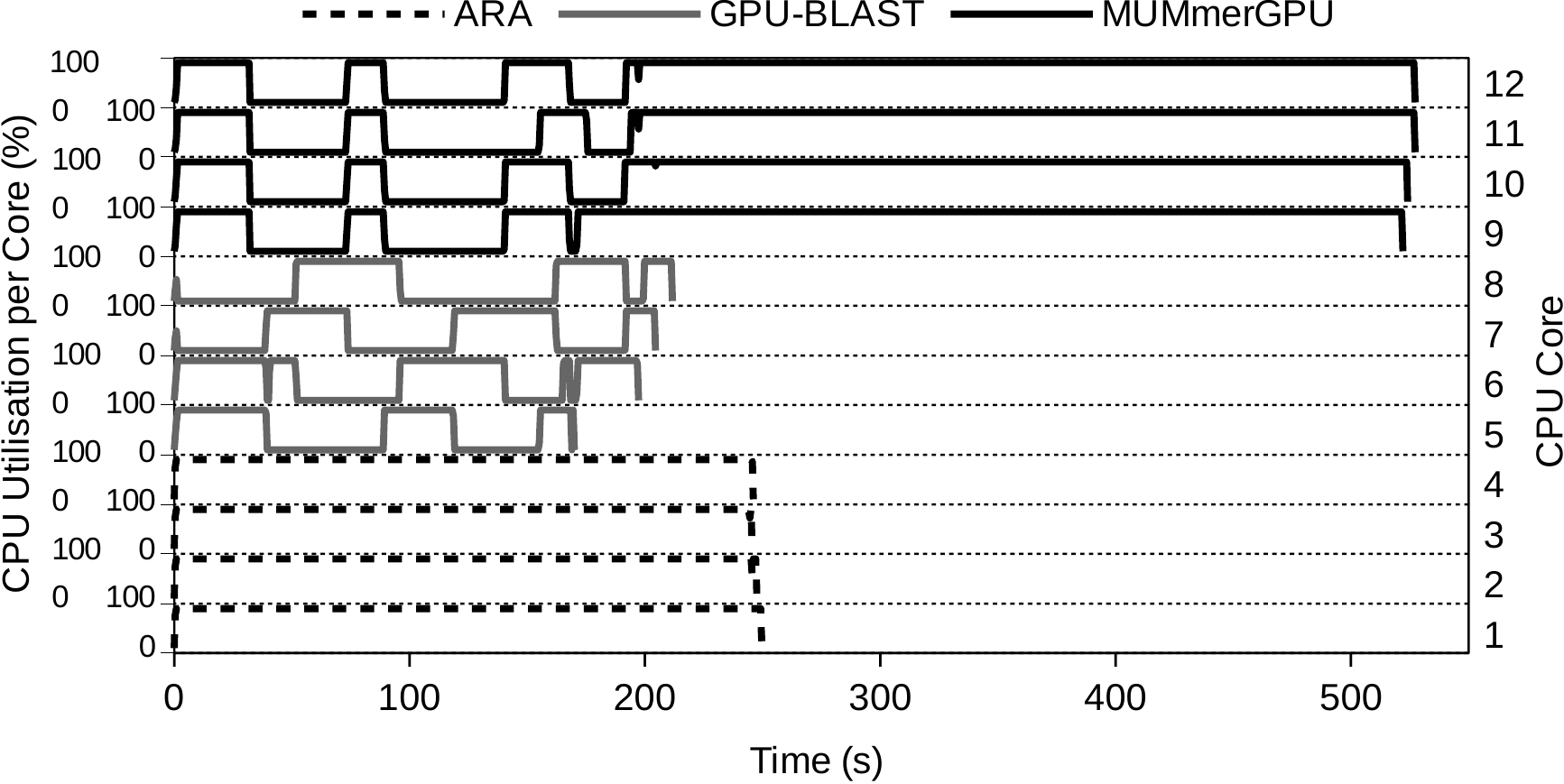}
}
\hfill
\subfloat[Using Priority FIFO policy]
{
	\label{fig:workload1_cpu_usage_PriorityFIFO}
	\includegraphics[width=0.52\textwidth]
	{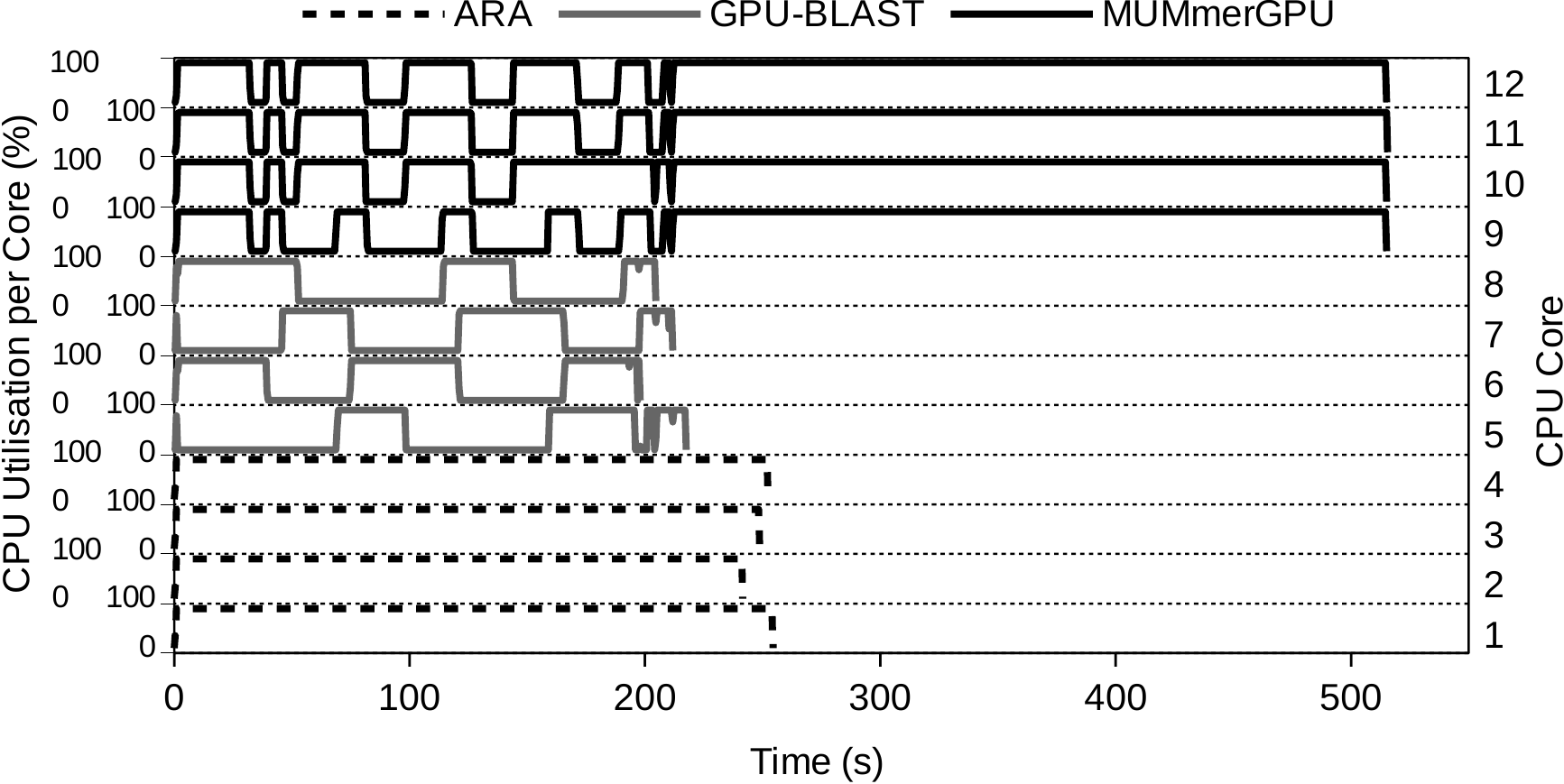}
} 
\hfill
\subfloat[Using Priority MMU policy]
{
	\label{fig:workload1_cpu_usage_PriorityMMU}
	\includegraphics[width=0.52\textwidth]
	{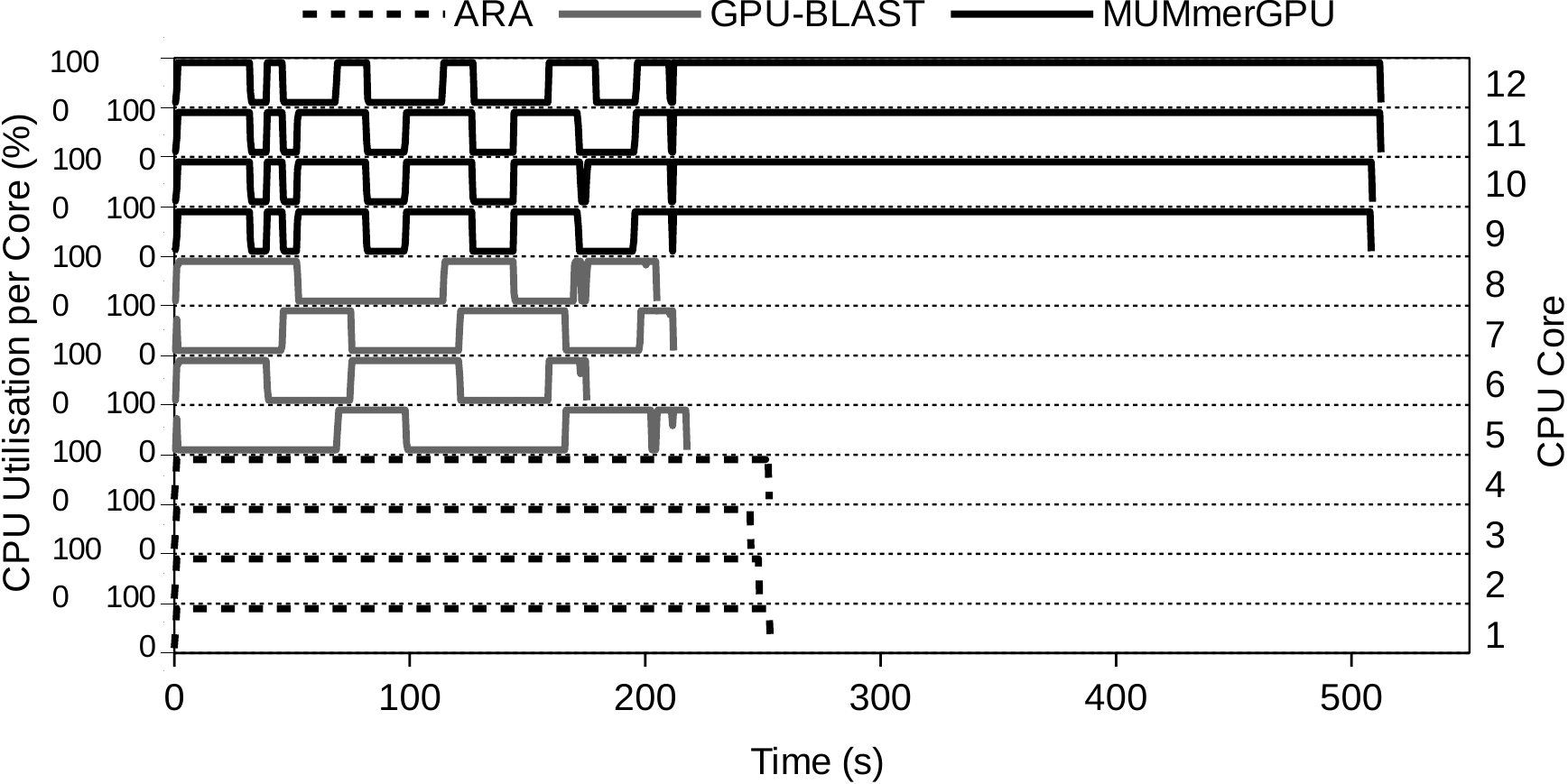}
}
\end{center}
\caption{Usage per CPU core when running a workload using schedGPU for different client notification policies.}
\label{fig:workload1_cpu_usage}
\end{figure}

Our experimental test-bed uses the Slurm~\cite{slurm} workload scheduler for scheduling jobs from multiple users. However, given that one GPU is used in the test-bed, Slurm handles multiple jobs requesting the GPU by sequentially executing them. As expected this results in the under-utilisation of the GPU. 

On the other hand, schedGPU can be employed to mitigate the above problem by managing the access of multiple job requests requiring GPUs. If there are $m$ real GPUs and $n$ CPUs, then Slurm is reconfigured (by only making changes to the configuration file) to be in possession of $m \times n$ GPUs. On our test-bed Slurm is reconfigured to have 12 GPUs (1 real GPU $\times$ 12 CPUs). This allows for Slurm to execute up to 12 concurrent jobs as if each CPU had access to a GPU. SchedGPU ensures that the jobs make use of the GPU safely. 

A workload comprising 12 concurrent jobs (12 jobs since there are 12 CPUs, each job requires one CPU and one GPU for execution) using 4 instances of ARA, MUMmerGPU and GPU-BLAST applications was submitted to Slurm. The applications have the same input as presented in the previous section. Figure~\ref{fig:workload1_cpu_gpu_usage} shows average CPU and GPU utilisation and average GPU memory utilisation for the workload. Figure~\ref{fig:workload1_cpu_usage} shows CPU utilisation of the cores for each application in the workload (Y-axis shows 0-100\% utilisation for each core). 

When considering non-priority based policies, it is observed that using FIFO (refer Figure~\ref{fig:workload1_cpu_gpu_usage_FIFO}) there are peaks in the GPU and GPU memory utilisation. This is because when sufficient memory is not available to furnish a request, no further requests are considered and hence memory remains under-utilised. However, using the MMU policy (Figure~\ref{fig:workload1_cpu_gpu_usage_MMU}) GPU memory utilisation is more evenly spread out. Requests of waiting clients are immediately furnished to maximise GPU memory usage. The MMU policy results in a reduction of nearly 3\% in the execution time of the workload over the FIFO policy as shown in Table~\ref{Table:gpu_usage_workload}. An improvement of nearly 1.5\% is also noted for both the average GPU utilisation and average GPU memory usage for MMU over FIFO. 

For non-priority based policies it is noted that CPU cores of the jobs waiting for GPU memory remain idle as shown in Figure~\ref{fig:workload1_cpu_usage_FIFO} and Figure~\ref{fig:workload1_cpu_usage_MMU}. This is noted for MUMmerGPU and GPU-BLAST instances since there is insufficient memory on the GPU to furnish all requests.

Since it is observed that the MUMmerGPU takes the most time for completing execution, all MUMmerGPU instances are assigned a high priority in an attempt to optimise the execution of the workload by further reducing the total execution time. When priority-based policies are taken into account, it is observed that the initial CPU utilisation increases and similar trends to non-priority based policies is observed for GPU utilisation.

In Figure~\ref{fig:workload1_cpu_usage_PriorityFIFO} and Figure~\ref{fig:workload1_cpu_usage_PriorityMMU}, assigning a higher priority to MUMmerGPU instances reduces waiting times for free GPU memory, thereby the CPU is idle for shorter periods of time. This translates into a reduction of total execution time using the priority-based policies by 15 seconds over the best case non-priority policy (MMU) as shown in Table~\ref{Table:gpu_usage_workload}. Similarly, an improvement of over 4.5\% and 9\% are noted for GPU utilisation and GPU memory utilisation, respectively, over the MMU policy.

%\textcolor{blue}{Finally, notice that the impact of the policies on performance and utilisation of both the GPU memory and GPU is rather small when compared to the naive FIFO policy. This is because the specific setup of the workload was selected to be generic and representative, so it does not bias the conclusions of this paper. Of course, there are scenarios where the priority policies might show more benefit, but as there is not space for showing different workloads, we have preferred to choose a generic one. Notice that even in this case, the priority policies presents some benefit.}

The advantage of using different policies on performance and utilisation is small when compared to the naive FIFO policy. This is because of the generic workload we have chosen in this paper to avoid a bias in our experimental results. Our workload comprises equal number of low, moderate and high GPU utilisation jobs. Even for such a workload there is some benefit in speeding up the overall execution time and utilisation. The benefit of the policies will be more apparent in workloads, for example, where a high memory utilisation job \textcolor{black}{blocks} a number of small memory utilisation jobs. A study on the effect of policies on different types of workloads is beyond the scope of this paper. 

\subsection{Summary}
%The experimental evaluation highlights that:
%\begin{enumerate}[leftmargin=0.3cm]
%\item[i.] The overhead of the shared memory approach is significantly less than that of the client-server approach, making it an ideal candidate for facilitating the schedGPU framework (refer Figure~\ref{fig:frameworks_comparison}). 

%\item[ii.] The performance gain, measured in terms of average speed-up, average GPU utilisation and average GPU memory utilisation, when concurrently executing individual applications using schedGPU is noted to be up to 10 times better than when not using schedGPU. 

%\item[iii.] For workloads comprising multiple applications, using Slurm in conjunction with schedGPU results in a speed-up of up to 5 times in the total execution time over the execution without schedGPU. The average GPU utilisation and average GPU memory utilisation is increased by 5 and 12 times, respectively, when compared to not using schedGPU.
%\end{enumerate}

We make three observations from the experiments. Firstly, the overhead of the shared memory approach is significantly less than that of the client-server approach, making it an ideal candidate for facilitating the schedGPU framework (refer Figure~\ref{fig:frameworks_comparison}). 

Secondly, the performance gain, measured in terms of average speed-up, average GPU utilisation and average GPU memory utilisation, when concurrently executing individual applications using schedGPU is noted to be up to 10 times better than when not using schedGPU. 

Thirdly, for workloads comprising multiple applications, using Slurm along with schedGPU results in a speed-up of up to 5 times in the total execution time. The average GPU utilisation and average GPU memory utilisation is increased by 5 and 12 times, respectively, when compared to not using schedGPU.

\begin{table}
\caption{Comparison of GPU utilisation and GPU memory utilisation when executing a workload comprising multiple applications}
\label{Table:gpu_usage_workload}
\begin{center}
%\begin{tabular}{ | l | p{1.25cm} | p{1.25cm} | p{1.25cm} | }
\begin{tabular}{ | l | r | r | r | }
\hline
                                    &          & \multicolumn{1}{|c|}{Average}		& \multicolumn{1}{|c|}{Average} \\
\multicolumn{1}{|c|}{Configuration}	& Time (s) & \multicolumn{1}{|c|}{GPU}			& \multicolumn{1}{|c|}{GPU} \\
                                    &          & \multicolumn{1}{|c|}{Utilisation}	& \multicolumn{1}{|c|}{Memory} \\
                                    &          & \multicolumn{1}{|c|}{(\%)}			& \multicolumn{1}{|c|}{Used (\%)} \\

\hline
Without schedGPU		& 2,485.20	& 9.24	& 3.79	\\ \hline
schedGPU FIFO			& 542.51	& 43.09	& 42.65	\\ \hline
schedGPU MMU			& 527.22	& 43.74	& 43.25	\\ \hline
schedGPU Priority FIFO	& 515.57	& 45.59	& 46.59	\\ \hline
schedGPU Priority MMU	& 512.53	& 45.75	& 47.15	\\ \hline
\end{tabular}
\end{center}
\vspace{-0.5cm}
\end{table}

%% file: relatedwork.tex
Approaches for efficiently utilising GPUs include (i) scheduling, (ii) kernel-based, (iii) synchronisation, and (iv) architectural approaches. 
Scheduling approaches include coarse-grain and fine-grain job scheduling. Coarse-grain job scheduling improves the overall throughput by scheduling concurrent applications on to nodes of the cluster~\cite{SchedulingRef1}. While throughput is improved, the focus is on inter-node scheduling of jobs, without considering a further level of optimisation at the intra-node level. Load balancing is commonly used for fine-grain job scheduling in multiple GPU environments~\cite{SchedulingRef3,SchedulingRef4}. However, the focus has been to uniformly distribute the executing workload across the GPUs, but not to improve utilisation of the GPUs. This paper focuses on intra-node scheduling at the fine-grain level to maximise GPU utilisation and to improve the overall throughput. 

Kernel-based approaches have included event-driven programming models for scheduling on shared GPUs~\cite{GLoop}. This approach does not concurrently share the GPU, but interleaves kernel executions on the GPU. A mechanism for concurrent execution of GPU kernels has been proposed~\cite{SchedulingRef5}. %However, current GPUs account for multiple context kernel concurrency. 
However, the mechanism does not safely handle GPU memory, such that sufficient GPU memory is available for the executing applications. A scheduler to facilitate multiple concurrent kernel executions has been proposed~\cite{KernelMerge}. Only two kernels can be executed and this may lead to potential deadlocks. More complex frameworks for synchronising GPUs have been developed~\cite{GPUSync,TimeGraph}. These require modifying the Linux kernel or GPU drivers, thereby limiting their use in production environments. %Similarly, an alternate real-time GPU scheduler has been developed by modifying the GPU drivers, only supporting a specific open-source driver~\cite{TimeGraph}.
Kernel-based approaches require extensive modifications, but the schedGPU framework requires \textcolor{black}{no modifications to the source code, if the implicit memory management functionality is used. As considered in Section~\ref{apifunctions}, the framework additionally provides explicit memory management functions requiring the source code to be minimally modified using the API, but offers the developer finer control over memory management. Our approach is simpler than modifying kernels.}
%carregon-v3\textcolor{red}{\sout{Our framework takes GPU memory into account and safely manages the access of applications to the GPU. The shared-memory approach is scalable for executing multiple instances of the same application as well as workloads with multiple applications since we do not use a centralised scheduler.}} 

Synchronisation approaches manage implicit and explicit synchronisations in GPU hardware and software for improving application concurrency~\cite{SchedulingRef6}. This approach avoids concurrent GPU operations to be executed sequentially. An application cannot use multiple kernel streams and cannot support unified memory. Our framework achieves synchronisation by a custom protocol we developed using file locks and system signals. %carregon-v3\textcolor{red}{\sout{The problem of abandonment that arises when other interprocess synchronisation mechanisms are used, such as mutexes, is mitigated.}}

Architectural approaches, such as Multi-Process Service (MPS)~\cite{CUDA-MPS-Paper, CUDA-MPS}
or Hyper-Q~\cite{CUDA-HyperQ-Paper, CUDA-HyperQ}, improves the GPU utilisation by allowing multiple processes, or threads, to simultaneously access a single GPU. However, in this research, GPU memory is not considered, and therefore, jobs fail when GPU memory is not available. The schedGPU framework on the other hand, safely handles GPU memory, and therefore applications do not fail due to insufficient memory but wait in a queue. 

Workload schedulers such as Torque~\cite{Torque}, PBS~\cite{PBS} or Slurm~\cite{slurm} include mechanisms for scheduling jobs on GPUs. We differentiate our framework from such schedulers in the following two ways. Firstly, schedulers operate at the cluster level (inter-node) and are capable of coarse-grain job scheduling, whereas schedGPU operates at the node level (intra-node) and performs fine-grain job scheduling to share the same physical GPU among multiple CPUs.
Secondly, the schedulers work ahead-of-time; the configurations need to be set before execution of the workload. However, schedGPU works just-in-time, such that scheduling is dynamic and occurs during the execution of the workload.
%carregon-v3\textcolor{red}{\sout{Moreover, it is noted that one of the merits of schedGPU is that jobs can be scheduled at the sub-millisecond timescales due to minimum overheads. Scheduling policies are incorporated in schedGPU for improving GPU and GPU memory utilisation.}}%, and providing the required quality of service to applications requesting preferential services.

%carregon-v3\textcolor{red}{\sout{There is prior research on job admission policies for schedulers based on GPU memory requests~\cite{rCUDA-Slurm-1, rCUDA-Slurm-2}. These are not generic solutions in that they require patching Slurm and can only be employed in a virtualised GPU environment. Our proposed research is a generic solution for any CUDA application.}}

Preemption mechanisms have been developed, but is based on including hardware extensions~\cite{preemption-1}. More recent GPU architectures provide preemption (for example, on the NVIDIA Pascal architecture)~\cite{TeslaP100}. Preemption prevents long-running applications that block other applications from monopolising the system. Such preemption mechanisms cannot inherently co-schedule applications. Our framework can be employed on both preemptible and non-preemptible GPUs and does not assume the GPU to be either time or space shared.
\textcolor{black}{A framework that can be used on non-preemptive accelerators to guarantee a given QoS for an application in terms of time duration by improving GPU utilisation is reported~\cite{Baymax}. Although QoS violation due to kernel interference and PCI-e bandwidth contention is minimised, it does not account for GPU memory-based co-scheduling.}

%% file: conclusions.tex
Currently, there are no schedulers that can safely co-schedule multiple GPU applications in terms of memory requirements. This results in the under-utilisation of GPUs in high-performance computing systems. In this paper, we aimed to improve the utilisation of GPUs by proposing an intra-node GPU scheduling framework, referred to as \textit{schedGPU}. We incorporated a client-server and shared memory approach for synchronising the access of multiple applications to the GPU. The schedGPU framework was validated using real-world applications both individually as single applications and collectively as workloads. A gain of over 10 times, as measured by performance speed-up, GPU utilisation and memory utilisation, was obtained for individual applications. For workloads, a speed-up of up to 5 times was noted and the average GPU utilisation and average GPU memory utilisation was increased by 5 and 12 times, respectively.

\textcolor{black}{
%\subsection{Future Work}
%\label{futurework}
\input{futurework}
}

%% file: futurework.tex
\textcolor{black}{
We intend to pursue the following three areas in our future research on schedGPU.
Firstly, exploring the performance of schedGPU against memory paging supported on new GPUs. We note that the memory paging mechanism is only available in the latest NVIDIA GPUs that use the Pascal architecture~\cite{TeslaP100}. Our approach, however, is also compatible with GPUs that do not employ memory paging (pre-Pascal); such GPUs are widely used in current HPC clusters and do not support memory paging. For example in the June 2017 Top500 list, NVIDIA GPUs used in clusters are all pre-Pascal GPUs. Using the NVLink high speed interconnect with Pascal GPUs may outperform schedGPU, but we expect competitive results when compared to the PCI-e version of Pascal GPUs. We anticipate that not all applications will benefit from GPU memory oversubscription and the page migration feature of the Pascal architecture. This is because memory access patterns are sometimes extremely difficult for prefetchers to predict. In these case, schedGPU may be used to complement memory paging for optimising application performance.}

%\textcolor{red}{
%\sout{Secondly, improving the programmability of schedGPU. The aim of this paper is to highlight the benefits of employing our approach for improving throughput. Programmers are required to minimally modify the GPU code by including schedGPU functions. However, modifying the source code can be avoided by developing a middleware which intercepts calls to CUDA. Then schedGPU operations are performed and the calls are forwarded to the actual CUDA function.}
%}

\textcolor{black}{
Secondly, by considering applications whose GPU memory requirement cannot be known before execution. It may not be always possible to know the total GPU memory required by an application as assumed in this paper. For example, when GPU memory is allocated at runtime; if two or more applications were concurrently executed and gradually increased their GPU memory usage, then when all GPU memory is used some of these applications could require more time to complete execution or may exit with a runtime error. In this case, schedGPU is not beneficial and other ways of improving performance will need to be explored.}

\textcolor{black}{
Finally, by accounting for applications that do not benefit from sharing the same GPU. For example, consider applications that require large amounts of GPU resources - many kernels, threads and register files per kernel. Even if GPU memory was available, it would not be beneficial to use schedGPU for co-scheduling another application due to the overheads in kernel switching. Running the applications exclusively and sequentially on the GPU may be beneficial. An application can be exclusively allocated to a GPU using schedGPU by simply pre-allocating all GPU memory to the application and then releasing it at the end of execution.}

%% file: acknowledgment.tex
This work was funded by Generalitat Valenciana under grant PROMETEO/2017/77.